\documentclass[12pt]{article}
\usepackage{amsmath,amssymb,amsfonts,amsthm,bbm}
\usepackage{epic,eepic,epsfig,longtable}
\usepackage{multirow,verbatim}
\usepackage{array}
\usepackage{graphicx}
\usepackage{mleftright}
\usepackage{xparse}
\usepackage{epsfig}
\usepackage{setspace}
\usepackage{CJK}
\usepackage{color}
\usepackage{enumitem}

\usepackage{algorithm}
\usepackage{algpseudocode}

\usepackage{lipsum}
\usepackage{mathtools}
\usepackage[authoryear,round]{natbib}
\usepackage{lineno,hyperref}
\usepackage{diagbox}

\usepackage{afterpage}
\usepackage{pdflscape}
\usepackage{stackengine}

\usepackage[normalem]{ulem}

\usepackage{booktabs}

\def\b#1{\textcolor{blue}{#1}}

\newcommand{\stkout}[1]{\ifmmode\text{\sout{\ensuremath{#1}}}\else\sout{#1}\fi}

\makeatletter
\def\singlespace{\def\baselinestretch{1}\@normalsize}

\numberwithin{equation}{section}

\renewcommand{\hat}{\widehat}

\renewcommand{\hat}{\widehat}

\newcommand{\bfm}[1]{\ensuremath{\mathbf{#1}}}

   \def\bA{\bfm A}  
   \def\bB{\bfm B}  
     
   \def\bD{\bfm D}  
     
  \def\bF{\bfm F}

   \def\bU{\bfm U}  
   \def\bV{\bfm V}  
\def\bw{\bfm w}     
     
   \def\bY{\bfm Y}

\newcommand{\bfsym}[1]{\ensuremath{\boldsymbol{#1}}}

              \def\bSigma{\bfsym \Sigma}
         
           \def\bOmega {\bfsym {\Omega}}

 \def\b1{\bfm 1}

\DeclareMathOperator{\argmax}{argmax}
\DeclareMathOperator{\argmin}{argmin}

\DeclareMathOperator{\cov}{cov}

\DeclareMathOperator{\diag}{diag}

\DeclareMathOperator{\var}{var}

\DeclareMathOperator{\tr}{tr}
\DeclarePairedDelimiter\abs{\lvert}{\rvert}
\DeclarePairedDelimiter\norm{\lVert}{\rVert}

\def\var{\mbox{var}}

\def\today{\ifcase\month\or
  January\or February\or March\or April\or May\or June\or
  July\or August\or September\or October\or November\or December\fi
  \space\number\day, \number\year}

\newdimen\biblioindent    \biblioindent=30pt

 at 8truept

\newcommand{\beq}{\begin{equation}}
  \newcommand{\eeq}{\end{equation}}
\newcommand{\beqn}{\begin{eqnarray}}
  \newcommand{\eeqn}{\end{eqnarray}}
\newcommand{\beqnn}{\begin{eqnarray*}}
  \newcommand{\eeqnn}{\end{eqnarray*}}

\allowdisplaybreaks
\usepackage{verbatim}
\renewcommand{\baselinestretch}{1.66}
\def\tilde{\widetilde}

\def\[{\left [}  \def\]{\right ]} \def\({\left (}  \def\){\right )}
 
\def\hat{\widehat}

\def \diag{\mathrm{diag}}

 \def \1 {\mathbf{1}}

\newtheorem{thm}{Theorem}

\newtheorem{lemma}{Lemma}

\theoremstyle{remark}
\newtheorem{remark}{Remark}
\theoremstyle{proposition}
\newtheorem{proposition}{Proposition}

\newtheorem{assumption}{Assumption}

\graphicspath{{figures/}{./images}}

\title{Property of Inverse Covariance Matrix-based Financial Adjacency Matrix for Detecting Local Groups} %

\author{Minseog Oh$^a$ and Donggyu Kim$^b$  \\
$^a$Korea Advanced Institute of Science and Technology (KAIST),\\
$^b$University of California, Riverside
}

\date{\today} %

\begin{document}

\maketitle %
\begin{spacing}{1.45}

\begin{abstract}
In financial applications, we often observe both global and local factors that are modeled by a multi-level factor model. 
When  detecting unknown local group memberships under such a model, employing a covariance matrix as an adjacency matrix for local group memberships is inadequate due to the predominant effect of global factors.
Thus, to detect a local group structure more effectively, this study introduces an inverse covariance matrix-based financial adjacency matrix (IFAM) that utilizes negative values of the inverse covariance matrix.
We show that IFAM ensures that the edge density between different groups vanishes, while that within the same group remains non-vanishing.
This reduces falsely detected connections and helps identify local group membership accurately. 
To estimate IFAM under the multi-level factor model, we introduce a factor-adjusted GLASSO estimator to address the prevalent global factor effect in the inverse covariance matrix.   
An empirical study using returns from international stocks across 20 financial markets demonstrates that incorporating IFAM effectively detects latent local groups, which helps improve the minimum variance portfolio allocation performance.

\end{abstract}


\section{Introduction}

Local factors, which influence only a specific subset of observed variables, have recently garnered increasing attention for their role in explaining certain economic and financial dynamics alongside global factors.
Empirical analyses have shown that local factors play a significant role in shaping results in economic and financial studies \citep{fama2012size, ferson1993risk, griffin2002fama, heston1994does, kose2003international}.
For example, local factors introduce heterogeneity into data due to their relationships and dynamics, which is often overlooked when focusing solely on global factors.

Based on this insight, several methodologies have been developed for analyzing multi-level factor models.
They usually stem from high-dimensional factor analysis and principal component analysis (PCA) \citep{ando2016panel, bai2015identification, choi2023large, choi2018multilevel, han2021shrinkage}.
One such method is the Double Principal Orthogonal complEment Thresholding (Double-POET) proposed by \citet{choi2023large}.
This method first applies PCA to estimate global factors.
It then applies PCA separately to each nationality-based group to estimate local factors from the remaining data after removing the global factors.
After both global and local factors are removed, it finally uses a thresholding procedure on what remains to capture the idiosyncratic volatility.
An important aspect of this procedure is the use of correct group memberships to ensure that variables are accurately associated with their respective local factors.
However, nominal group memberships, such as country, regional, or industry classification, may not fully capture the effects of local factors, which leads to potential estimation errors \citep{ando2017clustering}.
To address this issue, it is necessary to identify the latent group memberships that reflect the local factor effects.

A common approach for analyzing networks and detecting the group structure of assets is to construct an adjacency matrix.
In this matrix, each row and column correspond to a stock, and the entries represent the similarity between pairs of stocks.
Researchers often use the correlation of returns \citep{billio2012econometric, bonanno2004networks, chi2010network, diebold2014network, peralta2016network, vandewalle2001non}.
However, the correlation matrix is not suitable for detecting local groups in the presence of strong common factors because the dominant global factors can overshadow the local relationships between assets.
To remove the common factor effect, PCA is often employed; however, it can introduce a tuning parameter issue, such as incorrectly determining the number of global factors.
Even when the number of global factors is correctly identified, residual estimation errors may hinder the accurate detection of local groups.
Another way to analyze the asset network is to use precision matrices.
Precision matrices are conventionally employed in the context of Gaussian graphical models, where the off-diagonal entries with proper normalization using the diagonal elements of the matrix are interpreted as the negative conditional correlation between pairs of variables given the remaining variables \citep{fan2016overview,friedman2008sparse,ha2014partial,meinshausen2006high}.
However, to the best of our knowledge, no existing studies have explicitly linked the precision matrix to multi-level factor models and explored its property to distinguish local groups.

In this paper, we introduce the inverse covariance matrix-based financial adjacency matrix (IFAM) constructed by negative entries of an inverse covariance matrix to present the local group structure of asset returns.
The proposed IFAM highlights relationships between assets by reducing the influence of predominant common factors and amplifying the relatively smaller effects of local factors.
Specifically, the intuition behind focusing on the negative values of the inverse covariance matrix can be understood by considering its eigenvalue decomposition. 
Let the covariance matrix of asset returns, $\bSigma \in \mathbb{R}^{p\times p}$, be decomposed as $\bSigma = \bU \bV \bU^{\top}$, where $\bV_{1,1} > \cdots > \bV_{p,p} > 0$ are the eigenvalues. 
For some $k \in \mathbb{N}$, the matrix $\bU_{\cdot, 1:k}$ approximately spans the factor loading space and captures both global and local factors under pervasive conditions (see Assumption \ref{ass:sparsity}).
When two assets $i$ and $j$ belong to the same local group, they may share the same factor exposures.
Thus, the inner product of their factor loadings, $\bU_{i,1:k} \cdot \bU_{j,1:k}$, is expected to be positive.
However, since $\bU$ is orthonormal, the inner product of the entire loading vectors must be zero, that is,  $\bU_{i,\cdot} \cdot \bU_{j,\cdot} = 0$. 
As a consequence, the inner product of the residual factor loadings corresponding to the $k+1$ to $p$ dimensions, $\bU_{i,k+1:p} \cdot \bU_{j,k+1:p}$, must be negative to offset the positive contribution from the common local factor space.
This negative relationship is reflected in the inverse matrix, where the off-diagonal entries, $[\bSigma^{-1}]_{i,j} = \bU_{i,\cdot}^{\top} \bV^{-1} \bU_{j,\cdot}$, are more likely to be negative for pairs of assets within the same local group. 
Therefore, focusing on the negative values of the inverse matrix provides a natural way to detect local groups.
This finding aligns with the conventional Gaussian graphical model approach that uses negative partial correlations, which are calculated from a precision matrix, to identify groups \citep{bilgrau2020targeted, chandra2024functional,  fan2016overview, santiago2024comparing}.
Based on this intuition, we show the asymptotic property of IFAM.
Specifically, with a proper threshold level, the edge density within the same group remains non-vanishing asymptotically, while the edge density between different groups vanishes.
That is, connections within groups persist, while those between groups weaken.

In practice, since the true precision matrix is not observable, we need to estimate it. 
For a large number of assets, in the conventional Gaussian graphical model, 
precision matrix estimators, such as the graphical LASSO (GLASSO) \citep{friedman2008sparse} and the constrained $\ell_1$-minimization for inverse matrix estimation (CLIME) \citep{cai2011constrained}, require the sparsity condition. 
However, in a multi-level factor model, the presence of global factors makes it challenging to attain the required sparsity.
Hence, we propose a factor-adjusted GLASSO estimator.
This method first identifies global factors, then applies a sparse inverse matrix estimator, GLASSO, to the covariance after excluding the global factor component, and finally reconstructs the overall inverse covariance matrix using the Sherman-Morrison-Woodbury identity.
We demonstrate its asymptotic properties and achieve an elementwise $\ell_\infty$ convergence rate, which enables the factor-adjusted GLASSO estimator to construct IFAM with the same asymptotic properties as one constructed from the true inverse covariance matrix.
In practice, the clearer group separation property of IFAM leads to more accurate clustering results, which is essential for estimating local factor components.
These clustering results can be obtained using various clustering algorithms.
In our numerical studies, we employ the regularized spectral clustering (RSC) \citep{amini2013pseudo, chaudhuri2012spectral, joseph2016impact, qin2013regularized}.
Then, the detected labels are incorporated into the Double-POET procedure, which helps in more accurate estimation of the true large volatility matrix.
Furthermore, in the empirical study, we show that incorporating the proposed IFAM improves the performance of minimum variance portfolio allocations.

The rest of the paper is organized as follows.
In Section \ref{sec:IFAM}, we set up the model and propose IFAM and the factor-adjusted GLASSO estimator with their asymptotic properties.
In Section \ref{sec:simulation}, we conduct a simulation study to demonstrate the effectiveness of  IFAM in detecting local groups.
In Section \ref{sec:empirical}, we carry out an empirical study on portfolio allocation using the Double-POET method with group memberships identified by IFAM, which illustrates the practical advantage of using accurately identified group memberships.
Finally, we conclude in Section \ref{sec:conclusion}.
Appendix provides all proofs.

\section{Inverse covariance matrix-based financial adjacency matrix}\label{sec:IFAM}

\subsection{Model setup}\label{sec:model-setup}

We first fix some notations.
Let $\sigma_{\min}(\bA)$, $\sigma_{\max}(\bA)$, and $\sigma_i(\bA)$ denote the minimum, maximum, and $i$th largest eigenvalues, respectively, of the positive semi-definite matrix $\bA$.
Let $[\bA]_{i,j}$ denote the $(i,j)$th element of a matrix $\bA$.
Let $[\bA]_S$ represent the submatrix of $\bA$ formed by selecting rows and columns indexed by the elements of $S$, where $S$ is the ordered set of indices.

We consider the multi-level factor model as follows:
\begin{equation}\label{eq:base}
  y_{it} = b_{i}^{c\top} f^c_{t} + b_{i}^{g_{i}\top} f_{t}^{g_{i}} + u_{it}
 \quad \text{ for } i = 1, \ldots, p \quad \text{and} \quad t = 1, \ldots, T,
\end{equation}
where $y_{it}$ is the observed data for the $i$th asset at time $t$; $g_i$ is the group that includes the $i$th asset; $f^c_{t}$ and $b^c_{i}$ are $r_c \times 1$ vectors of latent global common factors and its corresponding factor loadings, respectively;
$f_{t}^{g_{i}}$ and $ b_{i}^{g_{i}}$ are $r_{g_{i}} \times 1$ vectors of latent local factors that only affect the group $g_{i}$ and its corresponding factor loadings;
and $u_{it}$ is an idiosyncratic error.
We consider the unknown group membership $g_i \in \mathcal{G} = \left\lbrace G_1,\ldots, G_J \right\rbrace$ and assume that the size of each group is $O(p^v)$ for some $v \in (0,1)$.
Therefore, we have $J=O(p^{1-v})$.
We can construct a concatenated factor loading matrix $\bB \in \mathbb{R}^{p \times r}$, where $r=r_c + \sum_{j=1}^{J} r_{G_j}$, $\bar{r}_{j}=r_c + \sum_{k=1}^{j} r_{G_k}$, and
\begin{eqnarray}\label{eq:concatenated-loading}
  [\bB]_{i,j}=
  \begin{cases}
  b^c_{i,j} , & \text{if } 1 \leq j \leq r_c , \cr
  b^{g_i}_{i,j-\bar{r}_{h(j)-1}} , & \text{if } g_i = G_{h(j)}, \text{ where }  h(j) = \min \left\lbrace k \in  \mathbb{N} |  j \leq \bar{r}_k  \right\rbrace, \cr
  0 , & \text{otherwise}.
  \end{cases}
\end{eqnarray}
Then, we can rewrite the model \eqref{eq:base} in vector form as
\begin{eqnarray*}
  Y_t &=& \bB_{c} f_t^{c} + \bB_{g} F_t^{g} + U_t \cr
  &=& \bB F_t + U_t ,
\end{eqnarray*}
where $\bB_{c} = [\bB]_{\cdot,1:r_c}$, $\bB_{g} = [\bB]_{\cdot,(r_c+1):r}$, $Y_t=\left( y_{1t},\ldots,y_{pt} \right) ^{\top}$, $F_t = \left( f_t^{c\top}, f_t^{G_1\top},f_t^{G_2\top},\ldots,f_t^{G_J\top} \right) ^{\top}$, $F_{t}^{g} = [F_{t}]_{(r_c+1):r}$, and $U_t = (u_{1t},\ldots,u_{pt})^{\top}$.
We can rewrite the model \eqref{eq:base} in matrix form as
\begin{eqnarray*}
  \bY &=& \bF_{c} \bB_{c}^{\top} + \bF_{g} \bB_{g}^{\top} + \bU \cr
  &=& \bF \bB^{\top} + \bU,
\end{eqnarray*}
where $\bY = (Y_{1},\ldots,Y_{T})^{\top} \in \mathbb{R}^{T\times p}$, $\bF=(F_1,\ldots,F_T)^{\top} \in \mathbb{R}^{T \times r}$, $\bU = (U_1,\ldots,U_T)^{\top} \in \mathbb{R}^{T \times p}$, $\bF_c = [\bF]_{\cdot,1:r_c}$, and $\bF_g = [\bF]_{\cdot,(r_c+1):r}$.
Then, the covariance matrix can be written as
\begin{eqnarray*}
  \bSigma &=&   \bB_{c} \bSigma_{c} \bB_{c}^{\top} + \bB_{g} \bSigma_{g} \bB_{g}^{\top} + \bSigma_u \cr
  &=&   \bB \bSigma_{F} \bB^{\top} + \bSigma_u ,
\end{eqnarray*}
where $\bSigma_{c} = \cov (f_t^{c})$, $\bSigma_{g} = \cov (F_t^{g})$, and $\bSigma_{F} = \cov (F_t)$.
We denote the precision matrix as $\bOmega = \bSigma^{-1}$.
We note that the identifiability condition for the latent factor model, such as the orthogonality, is not imposed. 
That is, the model \eqref{eq:base} is a general regression form.
On the other hand, within each local group, the loadings associated with each local factor retain the same sign to reflect the same direction of influence on that factor.
Without loss of generality, we assume that each loading is a unit vector (see Assumption \ref{ass:sparsity} (a)).

\subsection{Inverse covariance matrix-based financial adjacency matrix}\label{sec:IFAM-asymptotic}
The covariance matrix $\bSigma$ is not suitable as an adjacency matrix for detecting local groups because the strong influence of global factors, whose eigenvalues are of order $O(p)$, overshadows the smaller eigenvalues of local factors, which are of order $O(p^v)$.
To mitigate the predominant effect of global factors, we consider the inverse of the covariance matrix, $\bOmega$, since the inverse operation transforms the eigenvalues into their reciprocals, while the eigenvectors remain the same.
On the other hand, for the $i$th and $j$th assets that belong to the same group, the loadings corresponding to the first few eigenvalues may tend to have positive inner products because the effects of the corresponding local factors tend to be in the same direction. 
That is, the signs of the factor loadings in the same group tend to be the same.  
This implies that the loadings corresponding to the remaining eigenvalues tend to have negative inner products to satisfy the orthogonality of eigenvectors.
Thus, the off-diagonal entries of the inverse covariance matrix are more likely to be negative for pairs of assets within the same group.
Therefore, we define the inverse covariance matrix-based financial adjacency matrix (IFAM), $A_{\bOmega}$, as follows:
\begin{equation}\label{eq:IFAM}
  [A_{\bOmega}]_{i,j}
  =
  \begin{cases}
  -\bOmega_{i,j} , & \text{if } \bOmega_{i,j}<0 , \cr
  0 , & \text{otherwise.} \cr
  \end{cases}
\end{equation}
In this paper, we assume that the assets in the same local group have the same sign of the factor loadings.
That is, we consider that assets are in the same local group if their signs of the latent local factor loadings are the same.

For better understanding, we consider a simplified case with one global factor and one local factor for each group:
\begin{equation*} 
  y_{it} = b_{i}^{c} f^c_{t} + b_{i}^{g_{i}} f_{t}^{g_{i}} + u_{it}
 \quad \text{ for } i = 1, \ldots, p \quad \text{and} \quad t = 1, \ldots, T.
\end{equation*}
In the empirical study, the strongest factor is usually the market factor, which implies that $b_{i}^{c}$, $i=1,\ldots,p$, are market betas from the capital asset pricing model (CAPM) \citep{avellaneda2010statistical,connor1993test,laloux2000random,plerou2002random}. %
 Thus, the signs of $b_{i}^{c}$'s are positive. 
That is, the market factor represents the comovement of every asset.  
Similarly, since the local factor governs the comovement of the local group, the signs of $b_{i}^{g_{i}}$'s tend to be the same. 
In the latent factor model, under a pervasiveness condition, the eigenvector corresponding to the first eigenvalue approximates the global factor loadings, and the subsequent eigenvectors, from the second to the $(1+J)$th, approximately span the local factor loading space.
To provide clearer insight into the structure of the inverse matrix by obtaining a more manageable elementwise expression, we assume that the global and local factors, along with the idiosyncratic returns, are uncorrelated.
Under a normalization condition for factor identification, that is, the sum of squared loadings equals one for all factors, the inverse matrix can be written as
\begin{equation}\label{eq:simplified-inverse}
  [\bOmega]_{i,j} = \omega_{ij} - H_{ij} - Q_{ij} ,
\end{equation}
where $\omega_{ij} = [\bSigma_{u}^{-1}]_{i,j}$, $\sigma_{g_{i}}^{2} = \var (f_{t}^{g_i})$, $\sigma_{c}^{2} = \var (f_{t}^{c})$,
\begin{align*}
  & H_{ij} =
  \begin{cases}
  \omega_{ii} \omega_{jj} \frac{b_i^{g_i}  b_j^{g_i}  }{\sum_{k \in g_i} \omega_{kk} \left( {b_{k}^{g_i}} \right) ^2  +\sigma_{g_i}^{-2}} ,  & \text{if } g_i = g_j , \cr
  0 , & \text{otherwise} ,
  \end{cases} \cr
 & Q_{ij} = Q_3 (\omega_{ii} b_{i}^{c} - Q_{1,g_i} Q_{2,g_i} \omega_{ii} b_{i}^{g_i}) (\omega_{jj} b_{j}^{c} - Q_{1,g_j} Q_{2,g_j} \omega_{jj} b_{j}^{g_j}), \cr
 &  Q_{1,G} = \left( \sigma_{G}^{-2} + \sum_{k\in G} \omega_{kk} (b_{k}^{G})^{2}  \right)^{-1}, \quad 
  Q_{2,G} = \sum_{k \in G} \omega_{kk} b_{k}^{c} b_{k}^{G}, \cr
  &\text{and} \quad  Q_3 =  \left( \sigma_{c}^{-2} + \sum_{k=1}^{p} \omega_{kk}(b_{k}^{c})^{2}  + \sum_{G \in \mathcal{G}}  \frac{\left( \sum_{k\in G} \omega_{kk} b_k^{c} b_k^{G}   \right)^{2} }{\sigma_{G}^{-2} + \sum_{k\in G} \omega_{kk} (b_{k}^{G})^{2} }  \right) ^{-1}.
\end{align*}
The detailed calculations, which utilize the Sherman-Morrison-Woodbury matrix identity, are provided in Appendix \ref{sec:rank1-spec}.
Based on the multi-level factor structure with the pervasive and   incoherence conditions (see Assumption \ref{ass:sparsity}), we have $\sigma_{c}^{2} \asymp O(p)$, $\sigma_{g_i}^{2} \asymp O(p^{v})$, $\omega_{ii} \asymp O(1)$, $b_{i}^{c} \asymp O(p^{-1/2})$, and $b_{i}^{g_i} \asymp O(p^{-v/2})$ for all $1 \leq i \leq p$.
Then, we have $H_{ij} \asymp O(p^{-v})$ within groups and $Q_{ij} \asymp O(p^{-1})$ for any $i$ and $j$, which implies $Q_{ij}$ is negligible compared to $H_{ij}$ within groups.
We note that $H_{ij}$ and $Q_{ij}$ mainly come from the local and global factors, respectively.
By taking the inverse matrix, their magnitudes are changed as above.
Thus, we can successfully distinguish between the local and global factors.
Furthermore, since the signs of all $b_{i}^{g_{i}}$ are the same, $H_{ij}$ is positive. 
Consequently, under the sparsity condition of the inverse idiosyncratic volatility matrix $\bSigma_{u}^{-1}$, the off-diagonal entries of the inverse matrix are negative for pairs of assets within the same local group, while those between different groups are negligible.

It is worth noting that both IFAM and conventional partial correlation approach focus on the negative off-diagonal elements of the precision matrix to assess connections within groups.
Intuitively, from the perspective of the partial correlation approach, $\omega_{ij}$ primarily influences partial correlation, while $Q_{ij}$ and $H_{ij}$ can be interpreted as residual components coming from approximating global and local factors.
Unless $\omega_{ij}$ conveys additional local group information, this may lead to an ambiguous intuitive interpretation of the precision matrix for detecting local groups. 
In contrast, IFAM under the multi-level factor model highlights that $H_{ij}$ plays a central role in detecting local group structure. 
Specifically, the values of $H_{ij}$ are positive, larger in magnitude than $Q_{ij}$, and more prevalent within group relationships compared to the sparse inverse idiosyncratic volatility elements $\omega_{ij}$.
Therefore, $H_{ij}$ is a crucial component for accurately identifying local group structures.

To investigate the asymptotic behaviors of IFAM based on the true inverse, we need the following technical conditions.
\begin{assumption}\label{ass:sparsity}
  ~
  \begin{enumerate}[label=(\alph*)]
    \item \label{sparsity:normalized-pervasive} We have $\diag(\bB^{\top} \bB)=(1,\ldots,1)^{\top}$ and there exist positive constants $C_1$ and $C_2$ such that for any $1 \leq i \leq p$ and some $v \in (0,1)$,
    \begin{equation*}
      C_1 p^{-1/2} \leq b_i^c \leq C_2 p^{-1/2} \quad \text{and} \quad C_1 p^{-v/2} \leq b_i^{g_i} \leq C_2 p^{-v/2}
      .
    \end{equation*}

    \item \label{sparsity:factor-size} There exist positive constants $C_1, \ldots, C_6$ such that
    \begin{eqnarray*}
      && C_1 p \leq  \sigma_{i}(\bSigma_F)  \leq  C_2 p   \quad \text{for } i = 1, \ldots,  r_c , \cr
      && C_3 p^v \leq  \sigma_{i}(\bSigma_F)  \leq  C_4 p^v   \quad \text{for } i=r_c+1,\ldots, r,\cr 
      && C_5 \leq  \sigma_{i}(\bSigma_u) \leq  C_6   \quad \text{for } i = 1,\ldots, p.
    \end{eqnarray*}

    \item \label{sparsity:nontrivial-contribution} There exist positive constants $C_1$ and $C_2$ such that
    \begin{equation*}
      C_1 \leq \sigma_{\min} (\bB^{\top} \bB) \leq \sigma_{\max} (\bB^{\top} \bB) \leq C_2
      .
    \end{equation*}

    \item \label{sparsity:sparsity} There exist a constant $C$ and sparsity numbers $s_{b}=O(p^{\gamma_{s_{b}}})$ and $s_g=O(1)$ such that $\gamma_{s_b} < \min (\frac{v}{3},  \frac{1-v}{4} )$,
    \begin{align*}
      & \max_{G \in \mathcal{G}} \sum_{(i,j)\in G\times G^c} \b1([\bSigma_u^{-1}]_{i,j} \neq 0) \leq s_{b} ,\quad \max_{1 \leq i \leq p} \sum_{j \in g_i}  \b1([\bSigma_u^{-1}]_{i,j} \neq 0) \leq s_g , \text{ and }\cr
      &  \left\lVert \bSigma_u^{-1} \right\rVert _{\max}    < C
      .
    \end{align*}

  \end{enumerate}
\end{assumption}

\begin{remark}
  Assumptions \ref{ass:sparsity}\ref{sparsity:normalized-pervasive}--\ref{sparsity:factor-size} imply the pervasiveness of the factors that is essential to analyze latent factor models \citep{bai2003inferential, chamberlain1983arbitrage, choi2023large, fan2013large, fan2018ell, kim2019factor, lam2012factor, stock2002forecasting}. %
  The same signs on the loadings associated with each local or global factor are imposed to ensure that each factor reflects a uniform direction of influence.
  We only need the same sign condition. 
  In this study, for simplicity, we assume that the signs are positive. 
  Assumption \ref{ass:sparsity}\ref{sparsity:nontrivial-contribution} ensures that each global or local factor has a nontrivial influence \citep{bai2002determining,bai2003inferential,bai2009panel,stock2002forecasting}.
  Assumption \ref{ass:sparsity}\ref{sparsity:sparsity} is a sparsity condition for identifying local groups.
  The first and second sparsity conditions imply that there are sparse relationships between and within groups, respectively, in terms of idiosyncratic returns.
\end{remark}

The following theorem establishes the edge density within groups and between groups for IFAM.
\begin{thm}\label{prop:density}
  Under Assumption \ref{ass:sparsity}, for any group $G$, there exist positive values $\epsilon_{1,p} = o(p^{-v})$ and $\epsilon_{2,p} = O(s_g  p^{-v})$ such that
  \begin{equation}\label{eq:samegroup-density}
    \left|G \times G \right|^{-1}  \sum_{(i,j) \in G \times G} \b1 ([A_{\bOmega}]_{i,j} \geq  \delta_G |G|^{-1} - \epsilon_{1,p} ) >   \frac{\gamma_{G}^2 }{s_g^4(r_c+r_G)^{6}} - \epsilon_{2,p}
    ,
  \end{equation}
  where $s_b$ and $s_g$ are the parameters defined in Assumption \ref{ass:sparsity}, $\gamma_G = \frac{  \sigma^{2}_{\min}([\bSigma_{u}^{-1}]_{G}) }{16 \norm{[\bSigma_{u}^{-1}]_{G}}_{\max}^{2} C_G D}$, $D=\sigma_{\max}(\bB^{\top} \bSigma_{u}^{-1} \bB)$, $C_{G} = \left \lVert [\bB]_{G, \cdot} \right \rVert _{\max}^{2} |G|$, and $\delta_G = \frac{1}{4D} {\sigma_{\min}^{2}([\bSigma_{u}^{-1}]_{G})  \sqrt{r_c + r_G} }$  are constants not depending on $p$.
  Furthermore, for any group $G$, there exist positive values $\epsilon_{3,p} = o(p^{-v})$ and $\epsilon_{4,p} = O(s_b  p^{-1} )$ such that
  \begin{equation}\label{eq:diffgroup-density}
    \left|G \times G^c \right|  ^{-1} \sum_{(i,j) \in G \times G^{c}} \b1 ([A_{\bOmega}]_{i,j} \geq \epsilon_{3,p}) < \epsilon_{4,p} .
  \end{equation}
\end{thm}

Theorem \ref{prop:density} shows that the submatrix of the inverse covariance matrix, corresponding to the same group, contains an asymptotically non-vanishing ratio of significant negative values, while the submatrix corresponding to the between-group relationships has an asymptotically vanishing ratio of significant negative values.
Thus, IFAM can represent financial network connections by choosing an appropriate thresholding level of the order $O(p^{-v})$.  
This ensures that the groups can be effectively distinguished by clustering algorithms, such as minimum cut and spectral clustering \citep{coja2010graph,jerrum1998metropolis,mcsherry2001spectral,mossel2015reconstruction,rohe2011spectral}.
In practice, several approaches can be applied to choose an appropriate thresholding level.
One method involves selecting a threshold that ensures the thresholded graph exhibits desired topological properties, such as a specific edge density, average node degree, or network connectivity \citep{adamovich2022thresholding, langer2013problem, pan2023time, perkins2009threshold, zhou2018data}.
Another approach uses significance-based thresholding, where the thresholding level is set based on the statistical significance of edge weights under an assumed prior distribution \citep{ghoroghchian2021graph, ghosh2023selecting, roberts2017consistency, serrano2009extracting}.
This method ensures that only meaningful connections are retained in the graph.
Finally,  optimization-based selection chooses a thresholding level that maximizes an objective function.
In this approach, objective functions, often calculated through cross-validation, can include prediction accuracy or the silhouette index for clustering results, as well as criteria like minimizing the difference between the thresholded graph and the original weighted graph \citep{de2010inferring, dimitriadis2017topological, gates2014organizing, huang2012revealing}.

\subsection{Estimating IFAM from observed data}\label{sec:IFAM-observed}

Theorem \ref{prop:density} establishes the property of IFAM  based on the true inverse covariance matrix.
However, the true inverse covariance matrix is not available in practice, so it must be estimated from data.
In a high-dimensional setting, estimating the inverse covariance matrix directly from the algebraic inverse of the sample covariance matrix is challenging.
To tackle this issue, several estimators that utilize the sparsity structure of the inverse matrix, such as GLASSO \citep{friedman2008sparse} and CLIME \citep{cai2011constrained}, have been developed.
However, global factors make it hard to satisfy the necessary sparsity condition of the inverse covariance matrix.
To address this, we can use a factor-adjusted scheme and the Sherman-Morrison-Woodbury identity. 
For example, we first estimate the factor-adjusted input covariance matrix, $\bSigma_{E} = \bB_{g} \bSigma_{g} \bB_{g}^{\top} + \bSigma_u$, as follows:
$$
\hat{\bSigma}_{E} = \hat{\bSigma} - \sum_{i=1}^{r_c} \hat{\nu}_i \hat{\eta}_i \hat{\eta}_i^{\top},
$$
where  $\hat{\nu}_i$  and $\hat{\eta}_i$ are the $i$th largest eigenvalue of the sample covariance $\hat{\bSigma}$ and its corresponding eigenvector, respectively, and $r_c$ is the number of global factors.
Then, we apply the GLASSO method \citep{friedman2008sparse} to the factor-adjusted input covariance matrix to estimate $\bOmega_{E} = \bSigma_E^{-1}$ as follows:
\begin{eqnarray}\label{eq:FGLASSO-opt}
  \hat{\bOmega}_E = \argmax_{\bOmega \succ 0} \log \det \bOmega - \tr (\hat{\bSigma}_{E} \bOmega) - \rho_T \norm{\bOmega}_{1}
  ,
\end{eqnarray}
where $\rho_T$ is a tuning parameter determined in Proposition \ref{prop:GLASSO}. 
Finally, using the Sherman-Morrison-Woodbury identity,  the inverse covariance matrix is estimated as follows:
\begin{equation}\label{eq:factor-glasso}
  \hat{\bOmega} = \hat{\bOmega}_{E} - \hat{\bOmega}_{E} \hat{U} \left( \diag({\hat{V}})^{-1} + \hat{U}^{\top} \hat{\bOmega}_{E} \hat{U} \right)^{-1} \hat{U}^{\top} \hat{\bOmega}_{E}
  ,
\end{equation}
where  $\hat{V} \in \mathbb{R}^{r_c}$ and $\hat{U} \in \mathbb{R}^{p \times r_c}$ are the largest $r_c$ eigenvalues and corresponding eigenvectors for the sample covariance matrix $\hat{\bSigma}$, respectively.
We call this the factor-adjusted GLASSO estimator.
Using it, we can construct the estimated IFAM, denoted by $A_{\hat{\bOmega}}$, by substituting $\bOmega$ in \eqref{eq:IFAM} with its estimator, $\hat{\bOmega}$.

To obtain the same asymptotic property for IFAM constructed from the factor-adjusted GLASSO estimator, it is necessary to bound the elementwise $\ell_{\infty}$ convergence rate of factor-adjusted GLASSO estimator by $o(p^{-v})$.
To establish this convergence rate, the following conditions are required \citep{ravikumar2011high}.
\begin{assumption}\label{ass:GLASSO}    
~
  \begin{enumerate}[label=(\alph*)]
    \item There exist positive constants $a_1$, $a_2$, and $v_{*}$ such that for any $1 \leq i,j \leq p$ and $x \in (0,1/v_*]$, we have
    \begin{equation*}
      \mathbb{P}\left( \left| \hat{\bSigma}_{ij} - \bSigma_{ij} \right| \geq x  \right) \leq e^{-a_2 Tx^{a_1}}
      .
    \end{equation*}
    
    \item There exist a positive integer $d=O(p^v)$ such that
    \begin{equation*}
      \max_{1 \leq j \leq p} \sum_{i=1}^{p} \b1 (\bOmega_{E,ij} \neq 0) \leq d
      .
    \end{equation*}
    
    \item There exists some $\alpha \in (0,1]$ such that
    \begin{equation*}
      \max_{e \in S^c} \left\lVert \Gamma_{e,S} \left( \Gamma_{S} \right)^{-1}  \right\rVert _{1} \leq 1-\alpha
      ,
    \end{equation*}
    where $\Gamma = \bSigma_E \otimes \bSigma_E$ and $S = \left\lbrace (i,j) | 1 \leq i \neq j \leq p , \bOmega_{E,ij} \neq 0 \right\rbrace \cup \left\lbrace (1,1), \ldots, (p,p) \right\rbrace$.
    Here, $\Gamma_{e,S}$ denotes the submatrix of $\Gamma$ corresponding to the row indexed by $e$ and the columns indexed by $S$.
  \end{enumerate}
\end{assumption}
\begin{remark}
  Assumption \ref{ass:GLASSO}(a) implies the exponential-type tail condition.
  For instance, if returns follow a normal distribution, then we can set $a_1=2$.
  Assumption \ref{ass:GLASSO}(b) imposes the sparsity condition for the graph $\bOmega_{E}$, which represents group information.
  Assumption \ref{ass:GLASSO}(c) imposes an incoherence condition, which restricts the influence of the non-edge off-diagonal elements on the edge off-diagonal elements \citep{meinshausen2006high,ravikumar2011high,tropp2006just,wainwright2009sharp,zhao2006model}.
\end{remark}

The following proposition establishes the elementwise $\ell_{\infty}$ convergence rate for the factor-adjusted GLASSO estimator.
\begin{proposition}\label{prop:GLASSO}
  Under Assumptions \ref{ass:sparsity} and \ref{ass:GLASSO}, with $v<1/5$, let $c$ and $C$ be given positive constants.
  Then, there exists a positive constant $\bar{C}$ such that, for $T> \bar{C} p^{8v} \log p$ and $\rho_T=\alpha(48(2-\alpha) \kappa_{\Gamma}^{2}\kappa_{\Sigma}^3 d)^{-1}$, we have
  \begin{equation}\label{eq:GLASSO-rate}
    \left\lVert \hat{\bOmega} - {\bOmega} \right\rVert  _{\max} \leq \left\lVert \hat{\bOmega} - {\bOmega} \right\rVert  _{2} \leq C p^{-3v} + C p^{v} \sqrt{\frac{\log p}{T} }
  \end{equation}
  with probability at least $1-p^{-c}$, where $\kappa_{\Gamma} = \left\lVert \Gamma_{S} ^{-1} \right\rVert _{\infty}$ and $\kappa_{\Sigma} = \left\lVert \bSigma_E \right\rVert _{\infty}$.
\end{proposition}

Proposition \ref{prop:GLASSO} shows that the factor-adjusted GLASSO estimator achieves a convergence rate of $O(p^{v}(p^{-4v}+\sqrt{\frac{\log p}{T} }))$.
This convergence rate comes from two main components.
The term $O(p^{-4v}+\sqrt{\frac{\log p}{T} })$ originates from  estimating the inverse of the factor-adjusted covariance matrix using the GLASSO estimation procedure, and it is consistent with the results of Theorem 1 in \citet{ravikumar2011high}.
In contrast, the term $O(p^{v})$ comes from handling the Sherman-Morrison-Woodbury matrix identity. 
Specifically, the matrix multiplication of estimated matrices requires the $\ell_2$ convergence rate of the inverse estimation of the factor-adjusted matrix, which is discussed in Corollary 3 in \citet{ravikumar2011high}.
To consistently estimate the entire non-zero local group structure in the optimization problem in \eqref{eq:FGLASSO-opt}, we need a sufficient number of observations, such as $T> \bar{C} p^{8v} \log p$.

 In the following theorem, we extend the discussion to the case of the estimated inverse covariance matrix, $\hat{\bOmega}$, which can be obtained using the factor-adjusted GLASSO estimator.

\begin{thm}\label{prop:density-sample}
  Suppose that Assumption \ref{ass:sparsity} holds and
  \begin{equation}\label{eq:inv-est-cond}
    \left \lVert \hat{\bOmega} - \bOmega \right \rVert _{\max} = o(p^{-v}).
  \end{equation}%
  Then, for any group $G$, there exist positive values $\epsilon_{1,p} = o(p^{-v})$, $\epsilon_{2,p} = O(s_g  p^{-v})$ such that
  \begin{equation}\label{eq:samegroup-density-sample}
    \left|G \times G \right|^{-1}  \sum_{(i,j) \in G \times G} \b1 ([A_{\hat{\bOmega}}]_{i,j} \geq \delta_G |G|^{-1} - \epsilon_{1,p} ) > \frac{\gamma_{G}^{2} }{s_g^4 (r_c+r_G)^{6}} - \epsilon_{2,p}
    ,
  \end{equation}
  where $s_b$ and $s_g$ are the parameters defined in Assumption \ref{ass:sparsity}, $\gamma_G = \frac{  \sigma^{2}_{\min}([\bSigma_{u}^{-1}]_{G}) }{16 \norm{[\bSigma_{u}^{-1}]_{G}}_{\max}^{2} C_G D}$, $D=\sigma_{\max}(\bB^{\top} \bSigma_{u}^{-1} \bB)$, $C_{G} = \left \lVert [\bB]_{G, \cdot} \right \rVert _{\max}^{2} |G|$, and $\delta_G = \frac{1}{4D} {\sigma_{\min}^{2}([\bSigma_{u}^{-1}]_{G})  \sqrt{r_c + r_G} }$  are constants not depending on $p$.
  Furthermore, for any group $G$, there exist positive values $\epsilon_{3,p} = o(p^{-v})$ and $\epsilon_{4,p} = O(s_b p^{-1})$ such that
  \begin{equation}\label{eq:diffgroup-density-sample}
    \left|G \times G^c \right|  ^{-1} \sum_{(i,j) \in G \times G^{c}} \b1 ([A_{\hat{\bOmega}}]_{i,j} \geq \epsilon_{3,p}) < \epsilon_{4,p}    .
  \end{equation}
\end{thm}

Theorem \ref{prop:density-sample} shows that the estimated IFAM can obtain the same property as Theorem \ref{prop:density}. 
Furthermore, by Proposition \ref{prop:GLASSO}, for $T> \bar{C} p^{8v} \log p$,  the factor-adjusted GLASSO estimator has the elementwise $\ell_{\infty}$ convergence rate of $O(p^{-3v})$ with high probability, which is faster than the required rate in \eqref{eq:inv-est-cond}. 
Thus, the IFAM estimator constructed by the factor-adjusted GLASSO estimator can enjoy the properties of IFAM in Theorem \ref{prop:density-sample}.
We note that the results of Theorem \ref{prop:density-sample} hold as long as an inverse covariance matrix estimator satisfies \eqref{eq:inv-est-cond}.
The proposed factor-adjusted GLASSO estimator is one of them.

\subsection{Group identification via IFAM and its application to large volatility matrix estimation}\label{sec:IFAM-application}
The properties established in Theorems \ref{prop:density} and \ref{prop:density-sample} suggest that IFAM can be interpreted as a stochastic block model (SBM).
Several studies have demonstrated the inference procedures for SBM, such as the maximum likelihood estimator, variational estimator, and the use of spectral embedding, as well as their asymptotic properties \citep{allman2011parameter, ambroise2012new, celisse2012consistency, hagen1992new, lei2015consistency, mcsherry2001spectral,  rohe2011spectral, shi2000normalized}.
However, SBM has some potential issues, as thresholding always entails a loss of information, and the disruption of the latent structure becomes more severe when an incorrect threshold is selected, which lead to inaccurate clustering results \citep{aicher2015learning,thomas2011valued}.
In light of these challenges, other researchers have explored weighted adjacency matrix and studied its asymptotic properties, particularly with respect to spectral embedding techniques \citep{aicher2015learning,gallagher2024spectral,saade2014spectral,qin2013regularized,xu2020optimal,zhang2018understanding}.
To minimize the aforementioned problems, as well as the impact of tuning parameters, we employ the regularized spectral clustering (RSC) proposed by \citet{qin2013regularized} using IFAM in \eqref{eq:IFAM} as an input matrix to identify the group memberships of the financial assets.
Details for constructing the weighted adjacency matrix can be found in \eqref{eq:normalized-matrix}.
Under certain regularity conditions, the mis-clustering rate of the RSC method becomes asymptotically negligible as the sample size increases \citep{qin2013regularized}.
Details of the algorithm can be found in Appendix \ref{sec:clustering}.

In financial applications under the multi-level factor model, accurate local group labels are crucial to improve large volatility matrix estimation. %
In this study, we identify the local group labels using the RSC method with the proposed IFAM and incorporate these labels into the Double-POET procedure \citep{choi2023large}.
The procedure consists of three steps.
First, the global factor matrix is estimated by applying the PCA procedure to the sample covariance matrix.
Next, for each group identified using IFAM, we extract the submatrix of the sample covariance matrix that corresponds to that group after removing the global factors, and then apply the PCA procedure to estimate the local factor matrix.
The number of global and local factors are determined using the eigenvalue ratio method \citep{ahn2013eigenvalue}.
Finally, a thresholding scheme \citep{fan2013large, fan2016incorporating} is applied to the remaining residuals to estimate the idiosyncratic volatility matrix.
By summing the global, local, and idiosyncratic volatility matrices, the large volatility matrix is estimated.
Details of the Double-POET procedure can be found in \citet{choi2023large}.
The accurate group structure derived from IFAM enables more precise estimation of the local factor matrix.
As a result, the residual idiosyncratic component is also estimated more accurately.
Therefore, the accurate estimation of the local factor matrix leads to more precise estimation of both the idiosyncratic volatility and the resulting large volatility matrix.
In the empirical study, we find that incorporating IFAM shows the best performance in terms of minimum variance portfolio allocations. 

\section{Simulation study}\label{sec:simulation}

In this section, we conducted simulations to examine the properties of the proposed IFAM and evaluate its effectiveness in producing accurate clustering labels. %
We generated the data from the following multi-level factor model:%
\begin{equation}\label{eq:base-simulation}
  y_{it} = \sum_{l=1}^{5}  b_{il}^{c} f^c_{tl} + \sum_{l=1}^{2}  b_{il}^{g_i} f_{tl}^{g_i} + u_{it}
  \, \text{ for } i = 1, \ldots, p \quad \text{and} \quad t = 1, \ldots, T.
\end{equation}
The first factor loadings for the global and local factors, $b_{i1}^{c}$ and $b_{i1}^{g_{i}}$, respectively, were drawn from i.i.d. Uniform$(0.2, 1.8)$ and Uniform$(0.5, 1.5)$.
For the other factor loadings, $b_{il}^{c}$ and $b_{il}^{g_{i}}$ for $l \geq 2$, were generated as $b_{il}^{c} =  b_{i1}^{c} + U_{il}^{c}$ and $b_{il}^{g_{i}} =  b_{i1}^{g_{i}} + U_{il}^{g_{i}}$, where $U_{il}^{c}$ and $U_{il}^{g_{i}}$ were drawn from i.i.d. Uniform$(-0.16, 0.16)$ and Uniform$(-0.3, 0.3)$, respectively.
The global factors $(f^c_{t1},\ldots,f^c_{t5})^{\top}$ and local factors $(f^{g_{i}}_{t1},f^{g_{i}}_{t2})^{\top}$ follow $\mathcal{N}(0,\bSigma_{c})$ and $\mathcal{N}(0,\sigma_{g_{i}}^{2} \bSigma_{g})$, where $\sigma_{g_{i}}$'s were drawn from i.i.d. $\text{Gamma}(\alpha,\beta)$ with $\alpha=\beta=5$ for each $g_{i}$,
\begin{equation*}
  \bSigma_{c}
  =
  \begin{pmatrix}
    4.01  & -1 & -1 & -1 & -1 \cr
    -1 &  1 &  0    &  0    &  0    \cr
    -1 &  0    &  1 &  0    &  0    \cr
    -1 &  0    &  0    &  1 &  0    \cr
    -1 &  0    &  0    &  0    &  1 \cr
  \end{pmatrix},
  \text{ and }
  \bSigma_{g} = \begin{pmatrix} 0.26 & -0.25 \cr -0.25 & 0.25 \end{pmatrix}
  .
\end{equation*}
This data generating process (DGP) satisfies Assumption \ref{ass:sparsity}\ref{sparsity:normalized-pervasive}--\ref{sparsity:nontrivial-contribution}.

To set the sparse covariance matrix, we first generated $S=\diag(s_1,\ldots,s_p)$, where $s_i$'s were drawn from i.i.d. $\text{Gamma}(50,50)$.
Then, for each pair of different groups $j_1$ and $j_2$, we generated $d_{j_1} \in \mathbb{R}^{p}$ and $d_{j_2} \in \mathbb{R}^{p}$, where all elements of $d_{j_1}$ and $d_{j_2}$ are zeros except for one randomly chosen element in each group, with the nonzero value drawn from $\mathcal{N}(0,1/4)$.
We then set a sparse idiosyncratic covariance matrix as follows:
\begin{equation*}
  \bSigma_u = (0.03)^2 \left( S + \sum_{\substack{\left\lbrace j_1,j_2 \right\rbrace \subset \mathcal{G} \\ j_1 \neq j_2}} q_{j_1,j_2} (d_{j_1}+d_{j_2})(d_{j_1}+d_{j_2})^{\top}  \right) ^{-1}  \text{ and }  q_{j_1,j_2} \sim \text{Bern}(\frac{1}{|\mathcal{G}|\sqrt{\log |\mathcal{G}|}} )
  ,
\end{equation*}
where $\mathcal{G}$ is a set of groups.
This idiosyncratic volatility matrix satisfies the sparsity condition in Assumption \ref{ass:sparsity}\ref{sparsity:sparsity}.
We varied the sample size $T=250,500,1000$, the number of groups $|\mathcal{G}|=10,20,30$, and the number of assets in each group $|G|=10,20,30$.
We repeated the simulation 500 times for each setting.

To construct IFAM, we estimated the inverse covariance matrix using the factor-adjusted GLASSO estimator in \eqref{eq:factor-glasso}.
For the inverse of the factor-adjusted covariance, we used the GLASSO estimator \citep{friedman2008sparse} as follows:
\begin{eqnarray*}
  \hat{\bOmega}_E(\hat{\rho}) = \argmax_{\bOmega \succ 0} \log \det \bOmega - \tr (\hat{\bSigma}_{r_c^*,E} \bOmega) - \hat{\rho} \norm{\bOmega}_{1}
  ,
\end{eqnarray*}
where $\hat{\bSigma}_{r_c^*,E} = \hat{\bSigma} - \sum_{i=1}^{r_c^*} \hat{\nu}_i \hat{\eta}_i \hat{\eta}_i^{\top} $, $\hat{\nu}_i$ and $\hat{\eta}_i$ are the $i$th largest eigenvalue of the sample covariance $\hat{\bSigma}$ and its corresponding eigenvector, respectively, $r_c^*$ is the chosen number of common factors by the eigenvalue ratio method \citep{ahn2013eigenvalue}, and $\hat{\rho}$ is the tuning parameter determined by the BIC criterion \citep{lian2011shrinkage,wang2009shrinkage}:
\begin{equation*}
  \hat{\rho} = \argmin_{\rho} \text{BIC}(\rho) = \argmin_{\rho} \tr (\hat{\bSigma}_{r_c^*,E} \hat{\bOmega}_E(\rho))  - \log \det \hat{\bOmega}_E(\rho) + k_{\rho} \frac{\log T}{T} ,
\end{equation*}
where $k_{\rho}$ is the number of non-zero elements in the lower diagonal part of $\hat{\bOmega}_E(\rho)$.
We then applied the procedure in \eqref{eq:IFAM} to convert the estimated inverse matrix $\hat{\bOmega}$ into IFAM $A_{\hat{\bOmega}}$ and subsequently normalized the financial adjacency matrix as follows:
\begin{equation}\label{eq:normalized-matrix}
  \tilde{A}_{\hat{\bOmega}} = \bD_{\hat{\bOmega}}^{-1/2} {A}_{\hat{\bOmega}} \bD_{\hat{\bOmega}}^{-1/2}
  ,
\end{equation}
where $\bD_{\hat{\bOmega}} \in \mathbb{R}^{p \times p}$ is a diagonal matrix whose $i$th diagonal element is $[\hat{\bOmega}]_{i,i}$.

We first verified the property of IFAM proposed in Theorems \ref{prop:density} and \ref{prop:density-sample}, which implies that IFAM effectively limits the false positive rate (FPR) for detecting edges between groups while maintaining sufficient precision for detecting edges within groups.
With the thresholding level $\tau$, we can transform a normalized weighted adjacency matrix to a binary matrix as follows:
\begin{equation}\label{eq:binary-matrix}
\left( \b1 ([A_{ l}]_{i_1,i_2} > \tau) \right) _{i_1, i_2=1,\ldots, p},
\end{equation}
where $A_{l}$ is the normalized weighted adjacency matrix for the $l$th repeated simulation.  
A binary matrix is the form of observations in a stochastic block model.
For the $l$th repeated simulation and the group $G \in \mathcal{G}$, we calculated edge densities ($\text{ED}_{G,\tau,l}$) within and between groups varying $\tau$ from 0 to 1 as follows:
\begin{eqnarray} \label{eq-ED}
  &&\text{ED}_{G,\tau,l,\text{within}} = \frac{\sum_{(i_1,i_2) \in G \times G} \b1 ([A_{ l}]_{i_1,i_2} > \tau)}{|G \times G|} \qquad \text{and} \cr
  &&\text{ED}_{G,\tau,l,\text{between}} = \frac{\sum_{(i_1,i_2) \in G \times G^c} \b1 ([A_{ l}]_{i_1,i_2} > \tau)}{|G \times G^c|}
  .
\end{eqnarray}
When it comes to group membership identification, the minimum edge density within groups and the maximum edge density between groups are crucial \citep{abbe2018community, mossel2015reconstruction}.
Therefore, we calculated the mean edge density within and between groups as
\begin{eqnarray*}
  &&\text{ED}_{\tau,\text{within}}  = \frac{1}{500}  \sum_{l=1}^{500}  \min_{G \in \mathcal{G}} \text{ED}_{G,\tau,l,\text{within}} \quad \text{and} \cr
  &&\text{ED}_{\tau,\text{between}}  = \frac{1}{500} \sum_{l=1}^{500} \max_{G \in \mathcal{G}} \text{ED}_{G,\tau,l,\text{between}}.
\end{eqnarray*}
For comparison, we constructed a normalized weighted adjacency matrix based on the sample covariance matrix by adjusting the common factor effect as follows. 
We first calculated $\hat{\bSigma}_{r_c,u}=\hat{\bSigma} - \sum_{i=1}^{r_c} \hat{\nu}_i \hat{\eta}_i \hat{\eta}_i^{\top}$, and based on this matrix, constructed the (non-normalized) weighted adjacency matrix as follows:
\begin{equation*}
  A_{\hat{\bSigma}_{r_c,u}} =
  \begin{cases}
  |[\hat{\bSigma}_{r_c,u}]_{i_1,i_2}| , & \text{if } i_1 \neq i_2 , \cr
  0 , & \text{otherwise.} \cr
  \end{cases}
\end{equation*}
Then, we obtained the normalized weighted adjacency matrix, $\tilde{A}_{\hat{\bSigma}_{r_c,u}}$, by normalizing $A_{\hat{\bSigma}_{r_c,u}}$ using the normalization procedure \eqref{eq:normalized-matrix}. %
To obtain a binary matrix, we applied the same thresholding procedure described in \eqref{eq:binary-matrix} to the normalized weighted adjacency matrix.
To give more advantage to this method, we varied the number of common factors $r_c$ from one to ten and chose the one with the best performance.
We called it COV.

\begin{figure}[!h]
\centering
\includegraphics[width = 1\textwidth]{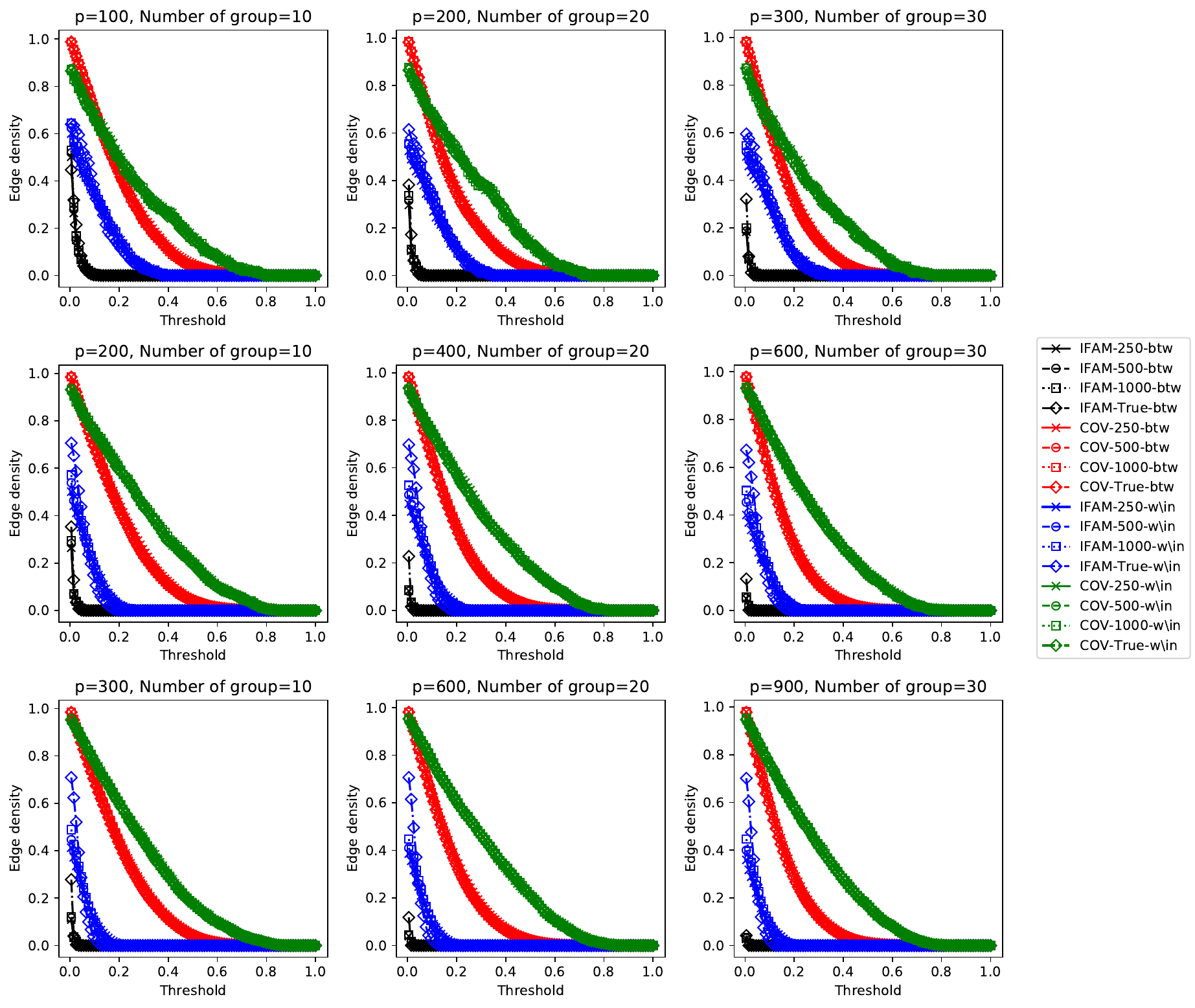}
\caption{The mean edge densities $ED_{\tau, \text{within}}$ and $ED_{\tau, \text{between}}$ for IFAM and COV adjacency matrices, where the edges were identified using the threshold $\tau$, varied from 0 to 1.
The adjacency matrices were constructed using sample sizes of 250, 500, and 1000, as well as the true covariance and inverse matrices.}
\label{Fig-positiveratio}
\end{figure}

\begin{figure}[!h]
\centering
\includegraphics[width = 0.8\textwidth]{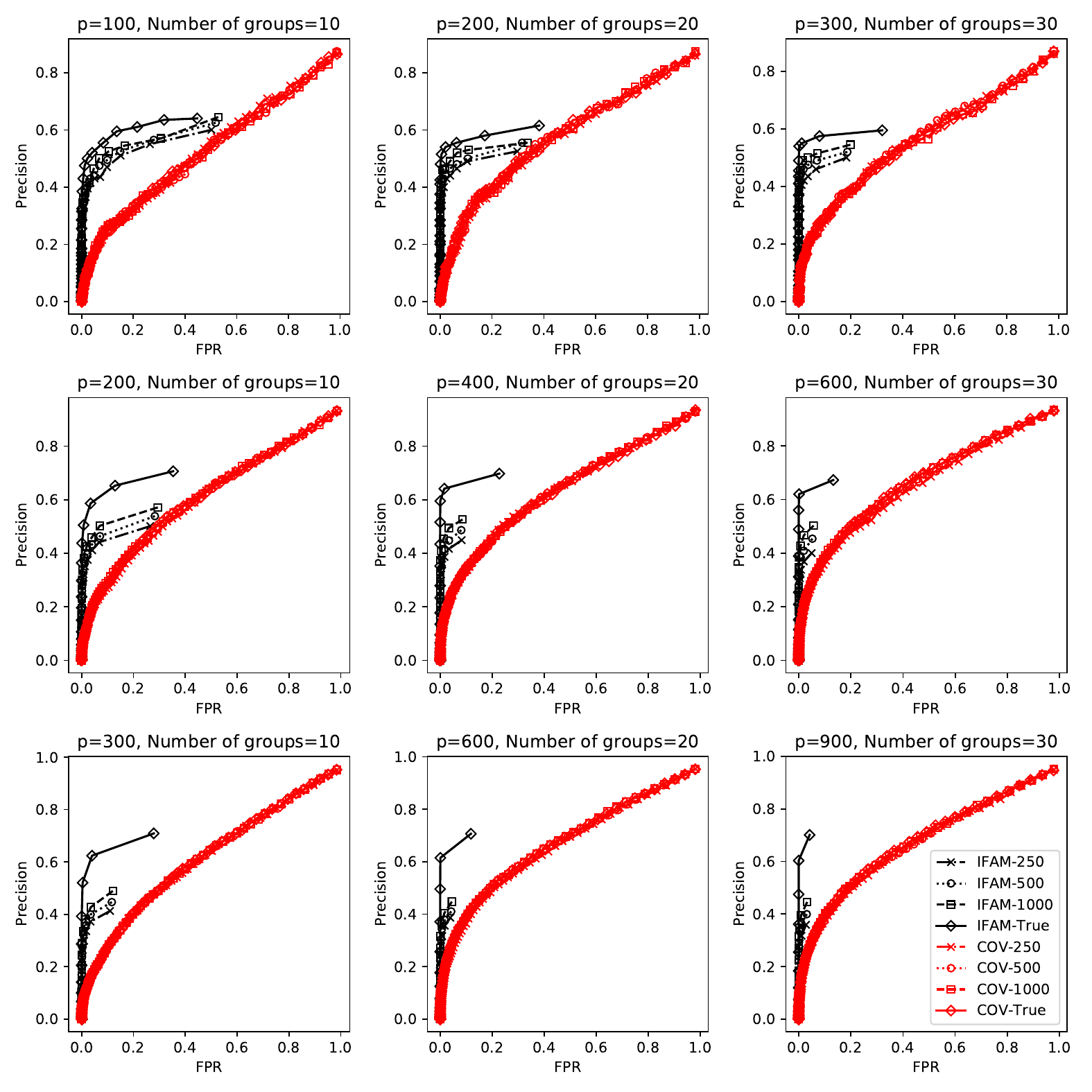}
\caption{Scatter plot of $ED_{\tau, \text{between}}$ (FPR) and $ED_{\tau, \text{within}}$ (precision) pairs for different values of the threshold $\tau$.
The adjacency matrices were constructed using sample sizes of 250, 500, and 1000, as well as the true covariance and inverse matrices.}
\label{Fig-FPR-Precision}
\end{figure}

Figure \ref{Fig-positiveratio} shows the edge densities $\text{ED}_{\tau, \text{within}}$ and $\text{ED}_{\tau, \text{between}}$ for IFAM and COV adjacency matrices, where the edges were identified using the threshold $\tau$ ranging from 0 to 1.
Figure \ref{Fig-FPR-Precision} displays the scatter plots of $\text{ED}_{\tau, \text{within}}$ and $\text{ED}_{\tau, \text{between}}$ pairs for different values of the threshold $\tau$.
We note that $\text{ED}_{\tau, \text{within}}$ and $\text{ED}_{\tau, \text{between}}$ correspond to precision and FPR, respectively.
The adjacency matrices were constructed using the sample sizes of 250, 500, and 1000, as well as the true covariance and inverse matrices.
For the COV adjacency matrix, we plotted the maximum mean edge density within groups and the minimum mean edge density between groups for $r=1,\ldots,10$.
From Figure \ref{Fig-positiveratio}, we find that the edge density between groups for IFAM vanishes drastically as the threshold $\tau$ increases, while for COV, it decreases more slowly.
From Figures \ref{Fig-positiveratio} and \ref{Fig-FPR-Precision}, we observe that IFAM maintains a higher edge density within groups compared to  COV at any given FPR level.
That is, the proposed IFAM can represent the group membership information well by choosing an appropriate thresholding level $\tau$. 
These findings are consistent with the results presented in Theorems \ref{prop:density} and \ref{prop:density-sample}.
When comparing the edge density within groups, the values of IFAM are consistently lower than those of COV.
Specifically, as the threshold $\tau$ approaches zero, COV's edge density within groups nears 1, while the edge density of IFAM is below some level. 
This is because IFAM removes non-negative values to reduce FPR.

We verified the effectiveness of the proposed IFAM in capturing the underlying group membership.
To achieve this, we applied the RSC algorithm \citep{qin2013regularized} to the constructed normalized weighted adjacency matrices to obtain the group membership $\phi_K$, where $K$ is the number of clusters. %
Details can be found in Appendix \ref{sec:clustering}.
To apply the clustering algorithm, we need to determine the number of clusters.
Since the Double-POET method can estimate the large volatility matrix accurately as long as the group membership is correctly identified \citep{choi2023large}, we chose the group membership $\phi_{\hat{K}}$ that minimizes the $C$-fold cross-validated log-likelihood loss of the Double-POET estimation as follows:
\begin{equation*}
  \hat{K} = \argmin_{K} \frac{1}{C}  \sum_{i=1}^{C} \text{tr}(\hat{\bSigma}_{-i} (\hat{\bSigma}_{i,\phi_{K}})^{-1}) -  \log \det ((\hat{\bSigma}_{i,\phi_K})^{-1})
  ,
\end{equation*}
where $K$ is the number of clusters, $\hat{\bSigma}_{-i}$ is the sample covariance matrix using the $i$th fold test samples, and $\hat{\bSigma}_{i,\phi_{K}}$ is the Double-POET estimation using the $i$th fold train samples with label $\phi_{K}$.
We set $C=2$.
To assess the accuracy of the clustering, we calculated the adjusted Rand index (ARI) \citep{hubert1985comparing} between the true group membership and the estimated group membership, $\phi_{\hat{K}}$.
Specifically, the ARI can be calculated as
\begin{equation*}
  \text{ARI} = \frac{\sum_{ij} \binom{n_{ij}}{2} - \left[ \sum_i \binom{a_i}{2} \sum_j \binom{b_j}{2} \right] / \binom{n}{2}}{\frac{1}{2} \left[ \sum_i \binom{a_i}{2} + \sum_j \binom{b_j}{2} \right] - \left[ \sum_i \binom{a_i}{2} \sum_j \binom{b_j}{2} \right] / \binom{n}{2}}
  ,
\end{equation*}
where $n_{ij}$ is the number of elements in both cluster $i$ in the true clustering and cluster $j$ in the predicted clustering, $a_i$ is the sum of elements in the true cluster $i$, $b_j$ is the sum of elements in the predicted cluster $j$, and $n$ is the total number of elements.
The ARI ranges from $-1$ to $1$, where $1$ indicates perfect clustering, and 0 represents random clustering.
The ARI is a well-known measure for evaluating clustering performance, as it adjusts for random chance, is not biased by the number of clusters, and remains fair even when cluster sizes are imbalanced \citep{hubert1985comparing, vinh2009information, warrens2022understanding}.

For comparison, we additionally considered the normalized weighted adjacency matrix based on GLASSO as well as the COV normalized weighted adjacency matrix.
Since off-diagonal elements of the normalized inverse covariance matrix can be interpreted as the partial correlation with reversed signs \citep{lauritzen1996graphical} in Gaussian graphical models, we multiplied the off-diagonal elements of the GLASSO estimation by $-1$ to make it a weighted adjacency matrix.
We also applied the same normalization as in \eqref{eq:normalized-matrix} to the weighted adjacency matrix.
For GLASSO, we treated this matrix as a signed weighted adjacency matrix \citep{chiang2012scalable, cucuringu2019sponge, cucuringu2021regularized, kunegis2010spectral} to align with the interpretation of partial correlations.
We applied the same clustering algorithm to the normalized weighted adjacency matrix to obtain the group membership from GLASSO.

\begin{figure}[!h]
\centering
\includegraphics[width = 0.8\textwidth]{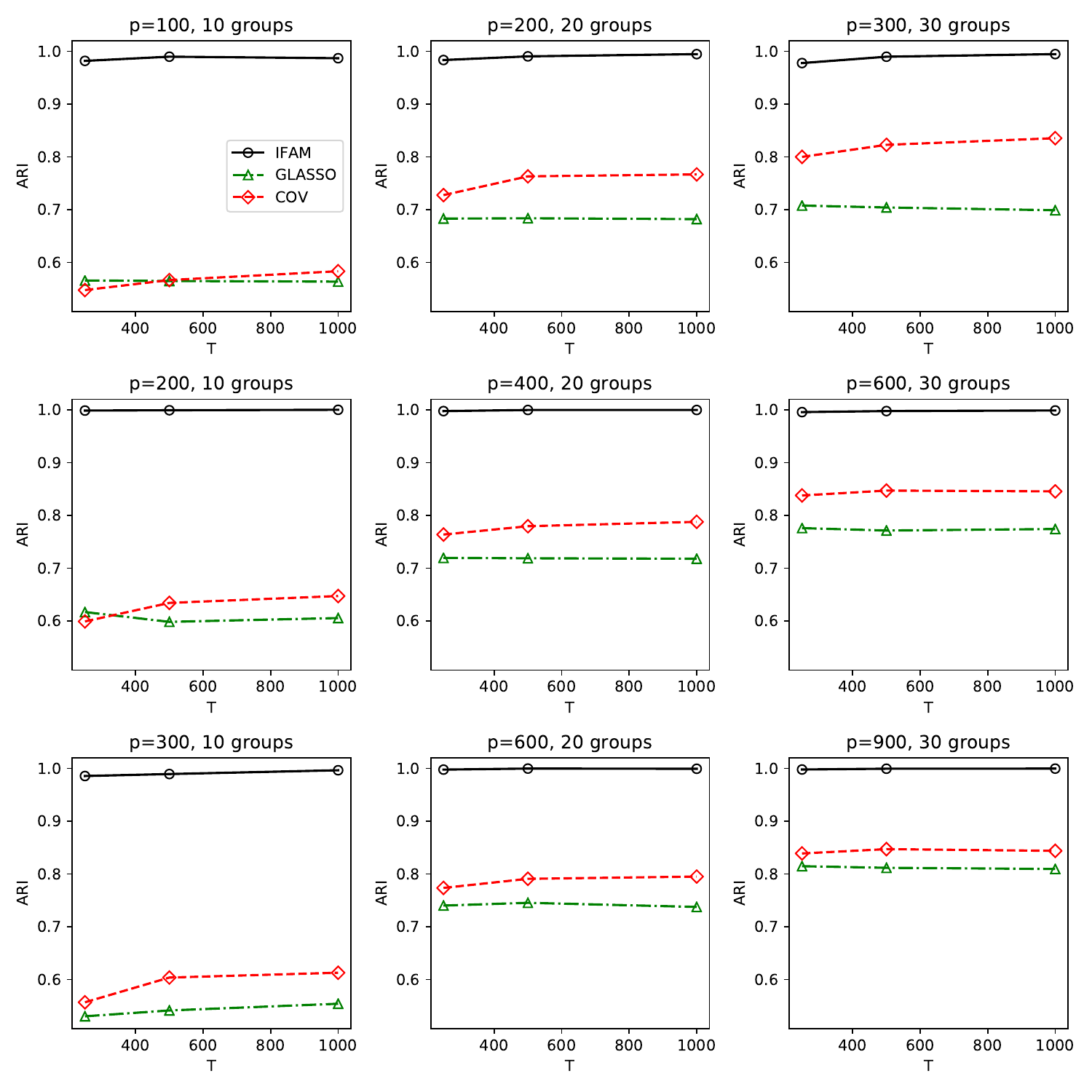}
\caption{Average ARIs for IFAM, COV, and GLASSO with the number of clusters set to 10, 20, and 30, and cluster sizes of 10, 20, and 30 for $T=250,500,1000$.} \label{Fig-ari} %
\end{figure}

Figure \ref{Fig-ari} shows average ARIs for IFAM, COV, and GLASSO with the number of clusters set to 10, 20, and 30, and cluster sizes of 10, 20, and 30, across varying $T=250,500,1000$.
From Figure \ref{Fig-ari}, we find that the average ARI of IFAM increases as the sample size $T$ increases.
This may be because IFAM can capture the underlying group membership more accurately as the sample size increases.
Furthermore, IFAM outperforms the other benchmarks in terms of average ARI across all settings.
It may be because IFAM effectively limits the false positive rate while maintaining sufficient precision for detecting edges, as shown in Figures \ref{Fig-positiveratio} and \ref{Fig-FPR-Precision}.

\begin{figure}[!h]
\centering
\includegraphics[width = 1\textwidth]{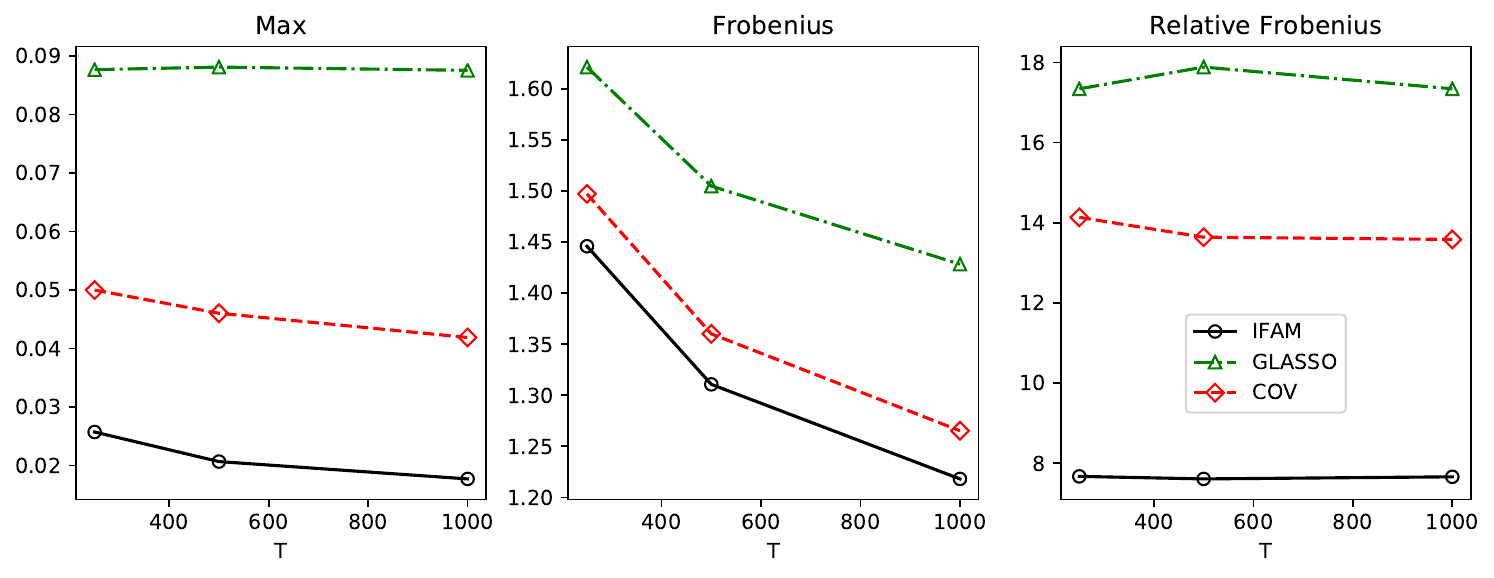}
\caption{Mean matrix max, Frobenius, and relative Frobenius norms for the matrix errors for the specification with $p=400$ and 20 groups.} \label{Fig-matrixerror}
\end{figure}

We assessed whether incorporating IFAM for group membership identification improves covariance matrix estimation.
To do this, we employed the Double-POET method \citep{choi2023large}, which requires group membership information.
The group membership was obtained from IFAM, GLASSO, or COV.
For the number of global and local factors, we used the eigenvalue ratio test \citep{ahn2013eigenvalue}.
To estimate idiosyncratic volatility, we followed the hard thresholding scheme in \citet{fan2013large}, which selects the minimum threshold value such that the thresholded matrix becomes positive definite for all values greater than this threshold.
We then calculated the mean matrix max, Frobenius, and relative Frobenius norms for the matrix errors to evaluate the accuracy of the estimated covariance matrix.
We display the results for the case where $p=400$ and there are 20 groups.
The remaining settings provided in Appendix \ref{sec:additional-simulation} and the results are similar. 
Figure \ref{Fig-matrixerror} shows the mean matrix max, Frobenius, and relative Frobenius norms for the matrix errors across varying $T=250,500,1000$. %
From Figure \ref{Fig-matrixerror}, we find that the mean matrix max, Frobenius, and relative Frobenius norms for the matrix errors decrease as the sample size $T$ increases.
Furthermore, IFAM outperforms COV and GLASSO in terms of the matrix max, Frobenius, and relative Frobenius norms for the matrix errors across all settings.
This is because IFAM effectively identified the group membership, which improves the large covariance estimation using the Double-POET method.

Finally, we checked the portfolio allocation performance using the Double-POET estimation with IFAM, GLASSO, and COV labels.
That is, given the estimated Double-POET volatility matrix, $\hat{\bSigma}_{\phi}$, where $\phi$ is the obtained group membership, we minimized the following portfolio risk function:
\begin{equation*}
\hat{\bw}_k = \underset{\bw_k \text{ s.t. } \bw_k^{\top} \mathbf{J}=1 \text{ and } \| \bw_k \|_1 \leq c_0 }{\text{argmin}} \bw_k^{\top} \hat{\bSigma}_{\phi} \bw_k,
\end{equation*}
where $\mathbf{J}=(1, \ldots, 1)^{\top} \in \mathbb{R}^p$ and $c_0$ is the gross exposure constraint that ranges from 1 to 4. 
We then computed the expected out-of-sample portfolio risk as $\hat{\bw}_k^{\top} \bSigma \hat{\bw}_k$, where $\bSigma$ is the true covariance.

\begin{figure}[!h]
\centering
\includegraphics[width = 1\textwidth]{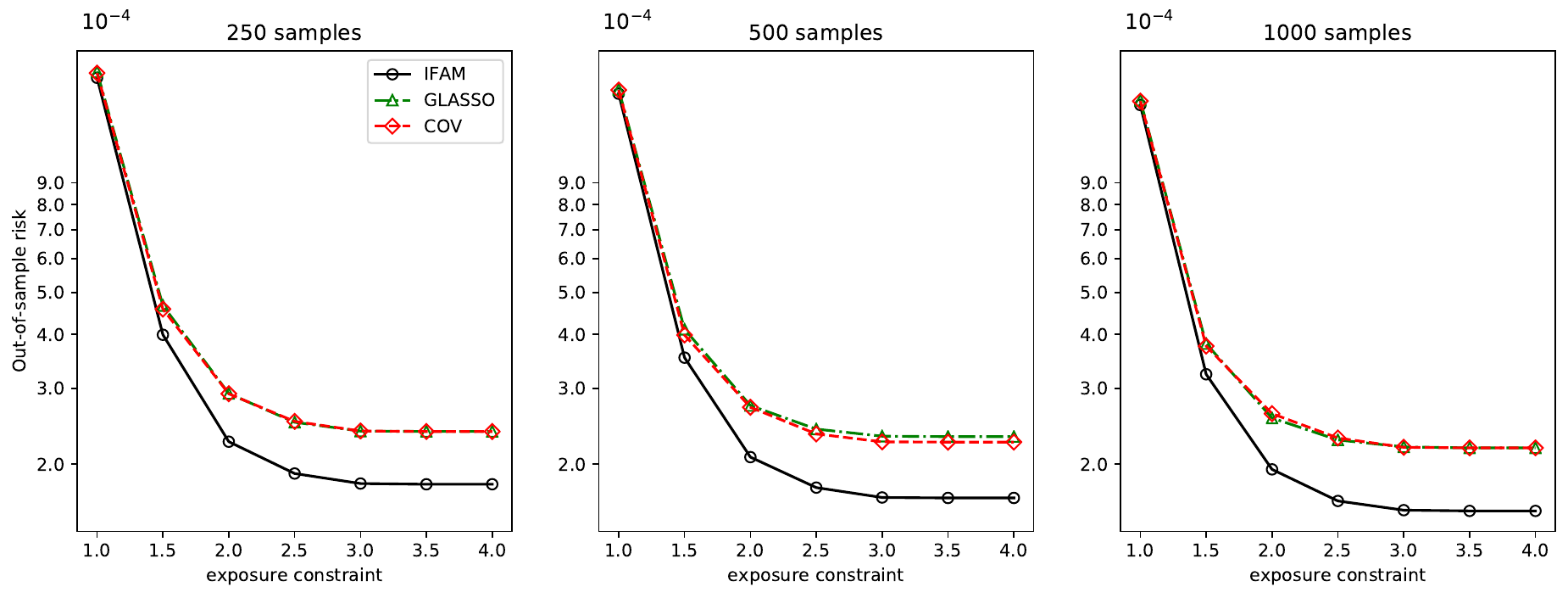}
\caption{Average expected out-of-sample portfolio risk using the Double-POET estimation with group memberships obtained from IFAM, GLASSO, and COV, against the gross exposure constraint $c_0$ for $T=250,500,1000$.}
\label{Fig-sim-risk}
\end{figure}

Figure \ref{Fig-sim-risk} shows the average expected out-of-sample portfolio risk using the Double-POET estimation with group memberships obtained from IFAM, GLASSO, and COV, against the gross exposure constraint $c_0$ for $T=250,500,1000$.
From Figure \ref{Fig-sim-risk}, we find that the average expected out-of-sample risk of IFAM decreases as the sample size increases, and using the group membership obtained from IFAM results in a lower average expected out-of-sample portfolio risk compared to using GLASSO and COV across all settings.
This is because the group membership obtained from IFAM is accurate, which enables the large volatility matrix to become more accurate as the sample size increases.
This results in better portfolio allocation performance.
From these results, we can conclude that IFAM effectively identifies the group membership, which improves the large covariance estimation and leads to better portfolio allocation performance.

\section{Empirical analysis}\label{sec:empirical}

In this section, we investigated the applicability of the proposed IFAM to empirical data.
We first applied the clustering algorithm to IFAM derived from the empirical data.
Then, we estimated the large volatility matrix using the Double-POET method based on the obtained group membership.
Finally, we conducted portfolio allocation using the estimated large volatility matrix.
We acquired weekly log-returns of global stock markets from January 2005 to December 2023 for 400 assets, evenly selected from 20 countries based on their total market capitalization, from the Compustat database in the Wharton Research Data Services (WRDS) system.
We used weekly returns to mitigate the effect of varying trading hours across countries.
The total number of sample weeks was 980.
We excluded stocks with missing data or no variation in this period.

\begin{figure}[h!]
\centering
\includegraphics[width = 1\textwidth]{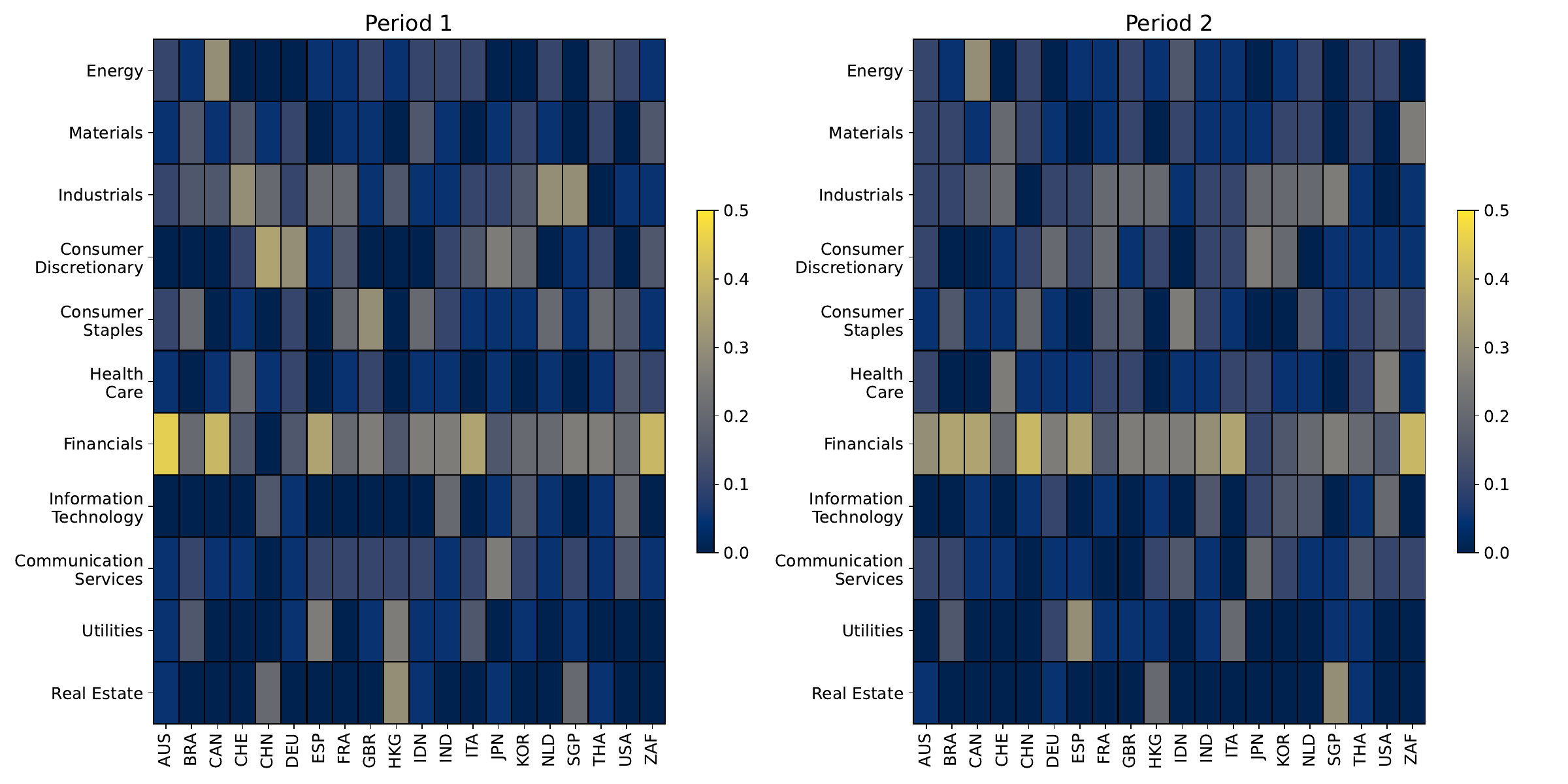}
\caption{The proportion of industry sectors across countries for two periods.} \label{fig:sector-proportion}
\end{figure}

We first examined the clustering results from the RSC method using IFAM.
To observe potential changes in the group structure over time, we considered two clustering results, one from January 2005 to December 2013 and the other from January 2014 to December 2023.
This division ensures a sufficient number of observations, 457 and 523, respectively, to estimate the inverse covariance matrix.

Figure \ref{fig:sector-proportion} shows the proportion of industry sectors by country in the dataset for both periods.
From Figure \ref{fig:sector-proportion}, we find that the proportion of the industry sector of each country is stable across the two periods, except for China.
In the case of China, the proportion of the financial sector increased while the proportion of the industrial sector decreased.

\begin{table}[h!]
\caption{Size of each group for two periods.}\label{tab:size}
\centering
\scalebox{0.85}{
\begin{tabular}{c c c c c c c c c c c c c c c c c c c c c}
\hline
Group & 1 & 2 & 3 & 4 & 5 & 6 & 7 & 8 & 9 & 10 & 11 & 12 & 13 & 14 & 15 & 16 & 17 & 18 & 19 & 20 \\
\hline
Period 1 & 93 & 42 & 28 & 27 & 23 & 22 & 20 & 20 & 20 & 20 & 16 & 15 & 14 & 13 & 11 & 10 & 6 &  &  &  \\
Period 2 & 59 & 29 & 26 & 22 & 20 & 20 & 19 & 19 & 19 & 18 & 18 & 18 & 17 & 17 & 16 & 16 & 15 & 14 & 10 & 8 \\
\hline
\end{tabular}
}
\end{table}

Table \ref{tab:size} reports the number of groups and the size of each group for the two periods.
Since the labeling of group membership was arbitrary, we renumbered the groups in descending order based on group size.
From Table \ref{tab:size}, we observe that the number of groups is smaller in the first period (17 groups) compared to the second period (20 groups).
Additionally, the largest group size in the first period is 93, while, in the second period, it is 59.
This suggests that the group structure in the second period may be more heterogeneous than in the first period, possibly due to the global financial crisis of 2008.
Specifically, during the first period, global stock markets were heavily influenced by the crisis and followed similar recovery patterns, which may have contributed to a more homogeneous group structure.
In contrast, the second period saw multiple geopolitical and economic disruptions, including the U.S.-China trade war, the COVID-19 pandemic, and the Russia-Ukraine war, which may have contributed to the more heterogeneous group structure.

\begin{figure}[!h]
\centering
\includegraphics[width = 1\textwidth]{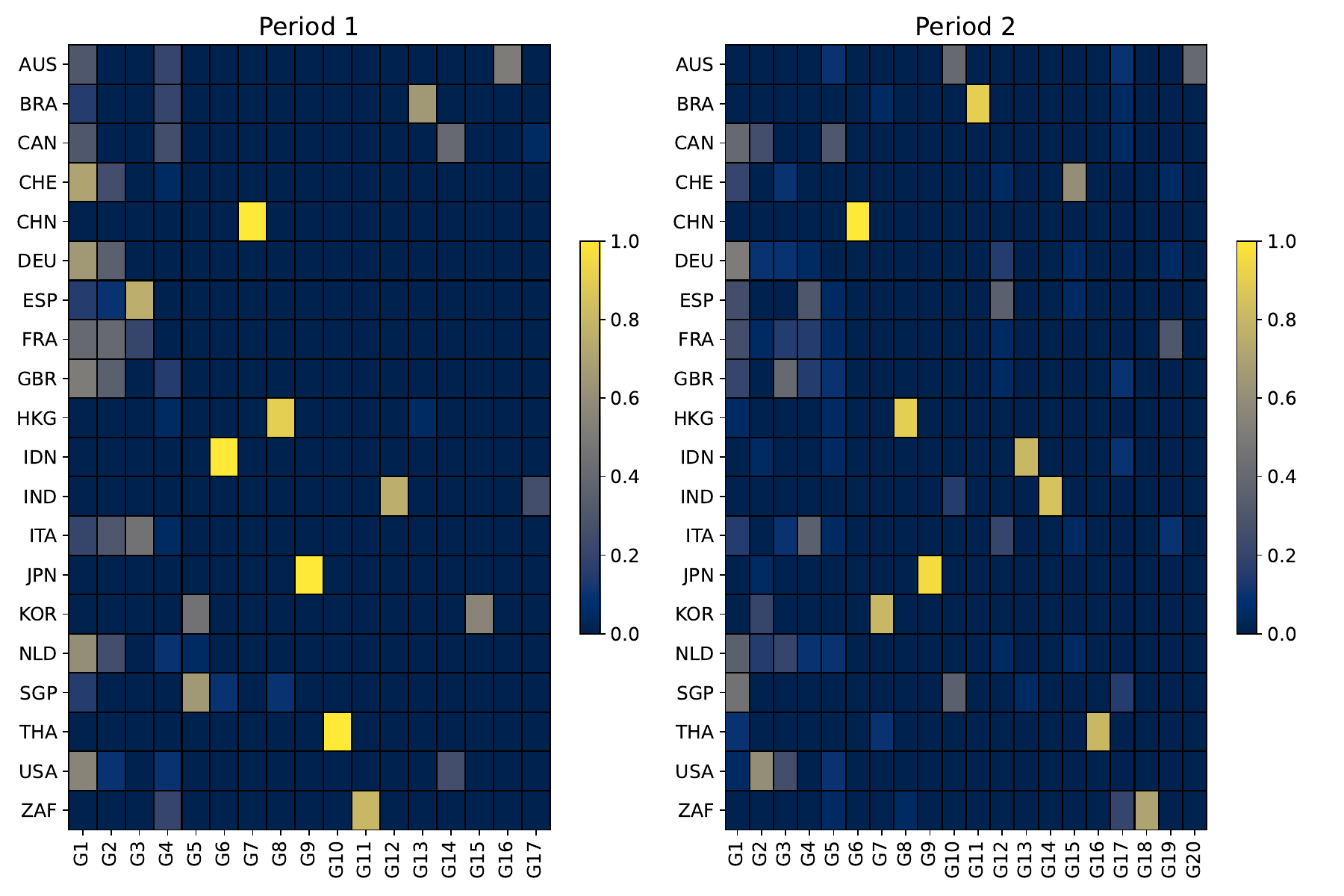}
\caption{Distribution of the firms in each country, where the $(i,j)$th element is the proportion of firms in a country $i$ such that they belong to $j$th group.} \label{fig:country-distribution}
\end{figure}

\begin{figure}[!h]
\centering
\includegraphics[width = 1\textwidth]{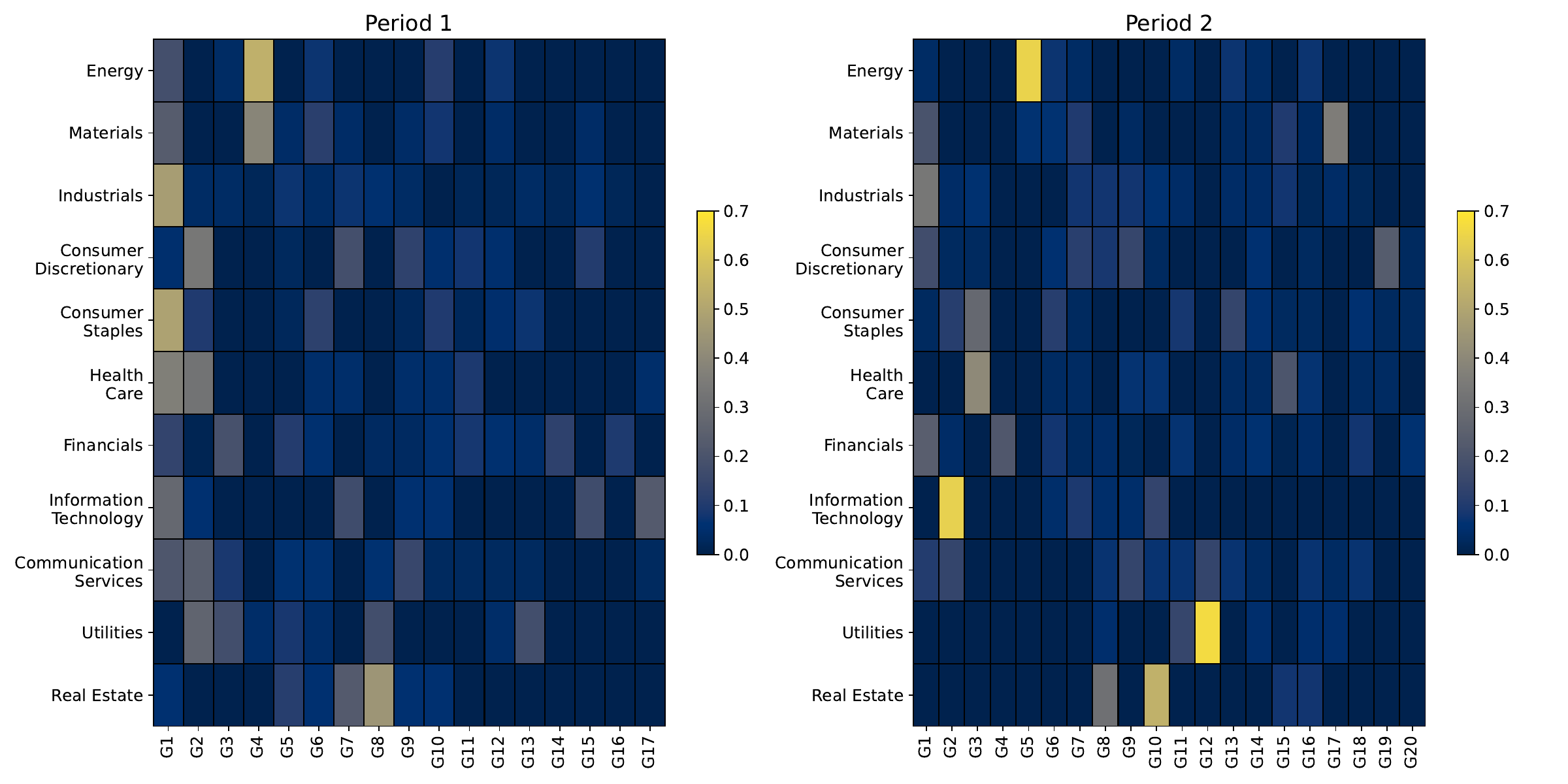}
\caption{Distribution of the firms in each sector, where the $(i,j)$th element is the proportion of firms in a sector $i$ such that they belong to $j$th group.} \label{fig:sector-distribution}
\end{figure}

To further understand how the groups are structured across countries and sectors, we checked the distribution of the firms in each sector and country for the two periods.
Figures \ref{fig:country-distribution} and \ref{fig:sector-distribution} show the distribution of the firms in each country and sector, respectively, for the two periods.
Specifically, the $(i,j)$th cell in the figures represents the proportion of firms in country or sector $i$ that belong to the $j$th group, $m_{ij}=n_{ij}/\sum_{k=1}^{J} n_{ik}$, where $J$ is the chosen number of groups and $n_{ik}$ is the number of firms that belong to country or sector $i$ and group $k$.
From Figure \ref{fig:country-distribution}, we observe that firms from each Asian country tend to form their own distinct group, and there are no significant changes in the distribution of firms within each country between the two periods.
From Figure \ref{fig:sector-distribution}, we find that in the first period, no sector is clustered within a specific group, while in the second period, several sectors, such as information technology, energy, and utilities, show a tendency to cluster within certain groups.
From these results, we can conclude that, in the first period, macroeconomic factors and country-level policies likely played a more dominant role, which led to less pronounced sectoral clustering.
In the second period, however, there appears to be an increased influence of sector-specific factors.
This influence is likely driven by global trends, such as technological advancements, energy transition, and regulatory changes in the utilities sector.

To investigate the economic benefit of IFAM, we investigated a minimum variance portfolio allocation performance using the group membership obtained from IFAM.
To estimate IFAM, we used 500 in-sample observations.
We estimated the large volatility matrix using the Double-POET method  with the obtained group memberships, $\phi$.
To reflect the current market conditions, we used the last 50 in-sample log-returns for the Double-POET estimation.
To estimate global and local factors, we followed the same procedure described in Section \ref{sec:simulation}.
For the thresholding of the idiosyncratic volatility part, we used the global industry classification standard (GICS). 
For example, we kept the volatilities within the same sector, but set others to zero \citep{fan2016incorporating}.
We then conducted the minimum variance portfolio allocation using the estimated volatility matrix, $\hat{\bSigma}_{\phi}$.
That is, given the estimated Double-POET volatility matrix, $\hat{\bSigma}_{\phi}$, we minimized the following portfolio risk function:
\begin{equation*}
\hat{\bw}_k = \underset{\bw_k \text{ s.t. } \bw_k^{\top} \mathbf{J}=1 \text{ and } \| \bw_k \|_1 \leq c_0 }{\text{argmin}} \bw_k^{\top} \hat{\bSigma}_{\phi} \bw_k,
\end{equation*}
where $\mathbf{J}=(1, \ldots, 1)^{\top} \in \mathbb{R}^p$ and $c_0$ is the gross exposure constraint that ranges from 1 to 8.
For a given period with $m$ weeks, we computed the annualized out-of-sample portfolio risk as
\begin{equation*}
  R = \sqrt{\frac{52}{m} \sum_{k=1}^{m} \left( \hat{\bw}_{k}^{\top} Y_{k+1} \right)^2   } ,
\end{equation*}
where $Y_{k+1}$ is the log-returns vector of the week $k+1$.
We applied the rolling window scheme and used two different out-of-sample periods--week 501 to week 740 and week 741 to week 980--and the whole out-of-sample period.

For comparison, we conducted portfolio allocation using other benchmark group memberships.
The first benchmark was based on nationality, which was not data-driven.
For the data-driven benchmarks, we utilized COV and GLASSO as inputs to the RSC algorithm.
These benchmark group memberships were then applied to the same Double-POET estimation procedure for portfolio allocation.
The results may depend on the specification of the number of global factors.
When we examined the number of global factors using the eigenvalue ratio test \citep{ahn2013eigenvalue} for every rolling window, the number of global factors varied from 1 to 5  with 1, 2, 3, 4, and 5 factors selected for 283, 182, 13, 1, and 1 cases, respectively.
Therefore, to assess whether the proposed group membership could outperform the benchmarks, we evaluated the results across this range of settings by varying the number of global factors from 1 to 5.
For each period and benchmark, we selected the specification that minimized out-of-sample portfolio risk.

\begin{figure}[!h]
\centering
\includegraphics[width = 1\textwidth]{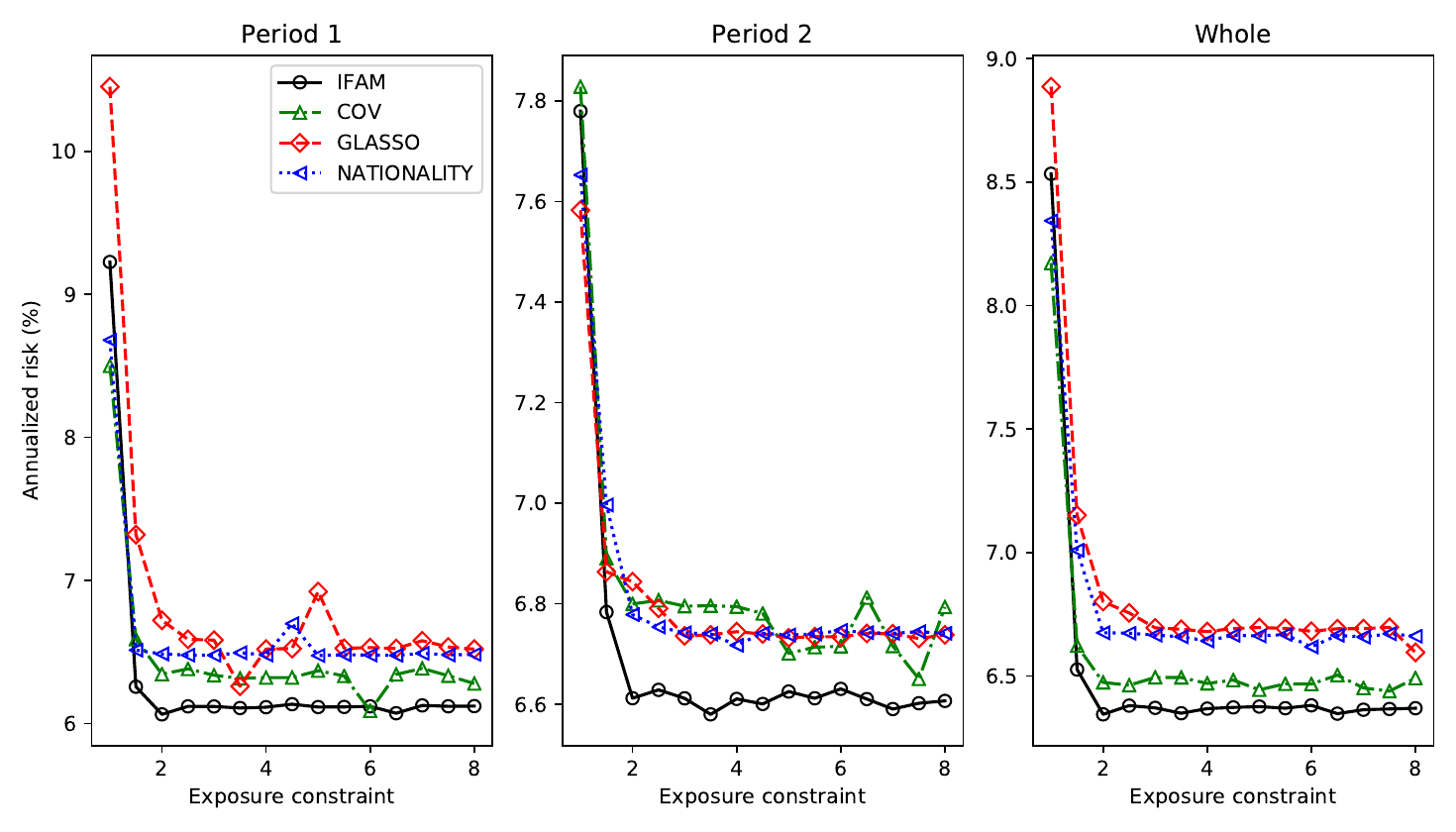}
\caption{Annualized out-of-sample portfolio risks using the Double-POET procedure with one of the group membership among IFAM, COV, GLASSO, and nationality for the two different periods and the whole period.}
\label{fig:empiric-risk}
\end{figure}

Figure \ref{fig:empiric-risk} plots the annualized out-of-sample portfolio risks obtained using the Double-POET procedure with different group memberships--IFAM, COV, GLASSO, and nationality--across two different periods as well as the whole period.
From Figure \ref{fig:empiric-risk}, we find that utilizing the proposed IFAM for the Double-POET procedure results in the lowest portfolio risk.
Furthermore, incorporating IFAM outperforms the other benchmarks under most exposure constraints.
This may be because the proposed IFAM effectively identifies local groups that share common local factors, which are crucial for accurately estimating the local factor component of the large volatility matrix.

\section{Conclusion}\label{sec:conclusion}
This study proposes a novel adjacency matrix for financial data that effectively identifies group membership under the multi-level factor model.
We show the asymptotic behaviors of the edge density within and between groups of IFAM.
Specifically, IFAM effectively reduces the false positive rate while it maintains sufficient precision for detecting edges.
To construct IFAM from data under the multi-level factor model, we propose the factor-adjusted GLASSO estimator.
We show that this method consistently estimates the inverse covariance matrix by addressing the prevalent global factor effect in the inverse covariance matrix.
In the empirical analysis, we also demonstrate that IFAM improves the minimum variance portfolio allocation performance.

\bibliography{sample}

\clearpage

\appendix

\section{Inverse matrix calculations under a simplified multi-level factor model}\label{sec:rank1-spec}
Let $[\bA]_{S_1,S_2}$ be the submatrix of $\bA$ formed by selecting rows indexed by the elements of $S_1$ and columns indexed by the elements of $S_2$, where $S_1$ and $S_2$ are the ordered set of indices.
Similarly, we define $[A]_{S} = (A_{i})_{i \in S}$ for a vector $A$ and a set of indices $S$.
We recall $\bSigma = \bB_{c} \bSigma_{c} \bB_{c}^{\top} + \bB_{g} \bSigma_{g} \bB_{g}^{\top} + \bSigma_u$ and $\bSigma_{E} = \bB_{g} \bSigma_{g} \bB_{g}^{\top} + \bSigma_u = \bOmega_{E} ^{-1}$.
Using the Sherman-Morrison-Woodbury matrix identity, we have
\begin{eqnarray*}
  \bOmega_{E} = \bOmega_{u} - \bOmega_{u} \bB_g S \bB_g^{\top} \bOmega_{u}
  ,
\end{eqnarray*}
where $S = (\bSigma_{g}^{-1} + \bB_{g}^{\top} \bOmega_{u} \bB_{g})^{-1}$.
Simple algebra shows that
\begin{equation*}
  [S]_{i,j} =
  \begin{cases}
  \frac{1}{\sum_{k \in g_i} \omega_{kk} \left( {b_{k}^{g_i}} \right) ^2+\sigma_{g_i}^{-2}} , & \text{if } i=j ,\cr
  0 , & \text{otherwise.}
  \end{cases}
\end{equation*}
Therefore, we have
\begin{eqnarray*}
  [\bOmega_{E}]_{i,j} = \begin{cases}
  \omega_{ii} - \omega_{ii}^{2} \frac{(b_i^{g_i}  )^2}{\sum_{k \in g_i} {\omega_{kk}} \left( {b_{k}^{g_i}} \right) ^2+\sigma_{g_i}^{-2}} , & \text{if } i=j ,\cr
  - \omega_{ii} \omega_{jj} \frac{b_i^{g_i}  b_j^{g_i}  }{\sum_{k \in g_i} {\omega_{kk}} \left( {b_{k}^{g_i}} \right) ^2+\sigma_{g_i}^{-2}} , & \text{if } i \neq j  \text{ and }  g_i = g_j ,\cr
  0, & \text{otherwise.}
  \end{cases}
\end{eqnarray*}
For any group $G$, we define a value $Q_{1,G} = 1/(\sigma_{G}^{-2} + \sum_{k\in G} \omega_{kk} (b_{k}^{G})^{2})$ and a vector $w_{G} = (\omega_{kk} b_{k}^{G})_{k\in G}$.
Then, for any $G \in \mathcal{G}$, we have
\begin{equation}\label{eq:omegaE-prop}
  [\bOmega_{E}]_{G} = [\bOmega_{u}]_{G} - Q_{1,G} w_{G} w_{G}^{\top}  \text{ and } [\bOmega_{E}]_{G,G^{c}} = 0 .
\end{equation}
We can rewrite \eqref{eq:omegaE-prop} as
\begin{equation}\label{eq:omegaE-wood-form}
  [\bOmega_E]_{i,j} = \omega_{ij} - H_{ij} ,
\end{equation}
where
\begin{align*}
  & H_{ij} =
  \begin{cases}
  \omega_{ii} \omega_{jj} \frac{b_i^{g_i}  b_j^{g_i}  }{\sum_{k \in g_i} \omega_{kk} \left( {b_{k}^{g_i}} \right) ^2  +\sigma_{g_i}^{-2}} ,  & \text{if } g_i = g_j , \cr
  0 , & \text{otherwise.}
  \end{cases}
\end{align*}
Using the Sherman-Morrison-Woodbury matrix identity, we have
\begin{equation}\label{eq:whole-woodbury}
  \bOmega = \bOmega_{E} - \bOmega_{E} \bB_c Q_3 \bB_c^{\top} \bOmega_{E} ,
\end{equation}
where $Q_3 = (\sigma_{c}^{-2} + \bB_{c}^{\top} \bOmega_{E} \bB_{c})^{-1}$.
Using \eqref{eq:omegaE-prop}, we have
\begin{eqnarray*}
  \bB_{c}^{\top} \bOmega_{E} \bB_{c} &=& \sum_{G \in \mathcal{G}}  [\bB_{c}]_{G}^{\top} [\bOmega_{E}]_{G} [\bB_{c}]_{G} \cr
  &=& \sum_{G \in \mathcal{G}}  [\bB_{c}]_{G}^{\top} ([\bOmega_u]_{G} - Q_{1,G} w_{G} w_{G}^{\top}) [\bB_{c}]_{G} \cr
  &=& \sum_{k=1}^{p} \omega_{kk}(b_{k}^{c})^{2}  + \sum_{G \in \mathcal{G}} Q_{1,G} ([\bB_{c}]_{G}^{\top} w_{G})^2  \cr
  &=& \sum_{k=1}^{p} \omega_{kk}(b_{k}^{c})^{2}  + \sum_{G \in \mathcal{G}}  \frac{\left( \sum_{k\in G} \omega_{kk} b_k^{c} b_k^{G}   \right)^{2} }{\sigma_{G}^{-2} + \sum_{k\in G} \omega_{kk} (b_{k}^{G})^{2} }  
  .
\end{eqnarray*}
Therefore, we have
\begin{eqnarray*}
  Q_3 =  \left( \sigma_{c}^{-2} + \sum_{k=1}^{p} \omega_{kk}(b_{k}^{c})^{2}  + \sum_{G \in \mathcal{G}}  \frac{\left( \sum_{k\in G} \omega_{kk} b_k^{c} b_k^{G}   \right)^{2} }{\sigma_{G}^{-2} + \sum_{k\in G} \omega_{kk} (b_{k}^{G})^{2} }  \right) ^{-1}
  .
\end{eqnarray*}
For any group $G$, we define a value $Q_{2,G} = \sum_{k \in G} \omega_{kk} b_{k}^{c} b_{k}^{G}$.
Simple algebra shows that
\begin{eqnarray*}
  [\bOmega_{E} \bB_{c}]_{G} &=& [\bOmega_{E}]_{G} [\bB_{c}]_{G} + [\bOmega_{E}]_{G,G^c} [\bB_{c}]_{G^c} \cr
  &=& [\bOmega_{u}]_G [\bB_{c}]_{G}  - Q_{1,G} w_{G}w_{G}^{\top} [\bB_{c}]_{G} \cr
  &=&  (\omega_{kk} b_{k}^{c} - Q_{1,G} Q_{2,G} \omega_{kk} b_{k}^{G})_{k\in G}
  ,
\end{eqnarray*}
where the second equality is due to \eqref{eq:omegaE-prop}.
Therefore, we have for any $1 \leq i,j \leq p$,
\begin{eqnarray*}
  [\bOmega_{E} \bB_c Q_3 \bB_c^{\top} \bOmega_{E}]_{ij} &=& 
  Q_3 [\bOmega_{E} \bB_c]_{i} [\bOmega_{E} \bB_c]_{j} \cr
  &=&  Q_3 (\omega_{ii} b_{i}^{c} - Q_{1,g_i} Q_{2,g_i} \omega_{ii} b_{i}^{g_i}) (\omega_{jj} b_{j}^{c} - Q_{1,g_j} Q_{2,g_j} \omega_{jj} b_{j}^{g_j})
  .
\end{eqnarray*}
Using \eqref{eq:whole-woodbury} and \eqref{eq:omegaE-wood-form}, we have
\begin{equation*}
  [\bOmega]_{i,j} = \omega_{ij} - H_{ij} - Q_{ij} ,
\end{equation*}
where $Q_{ij} = Q_3 (\omega_{ii} b_{i}^{c} - Q_{1,g_i} Q_{2,g_i} \omega_{ii} b_{i}^{g_i}) (\omega_{jj} b_{j}^{c} - Q_{1,g_j} Q_{2,g_j} \omega_{jj} b_{j}^{g_j})$.

\section{Additional simulation analyses}\label{sec:additional-simulation}

In this section, we present the full simulation results for all cases, including varying the number of clusters from 10 to 30 and cluster sizes from 10 to 30.
\begin{figure}[!h]
\centering
\includegraphics[width = 0.8\textwidth]{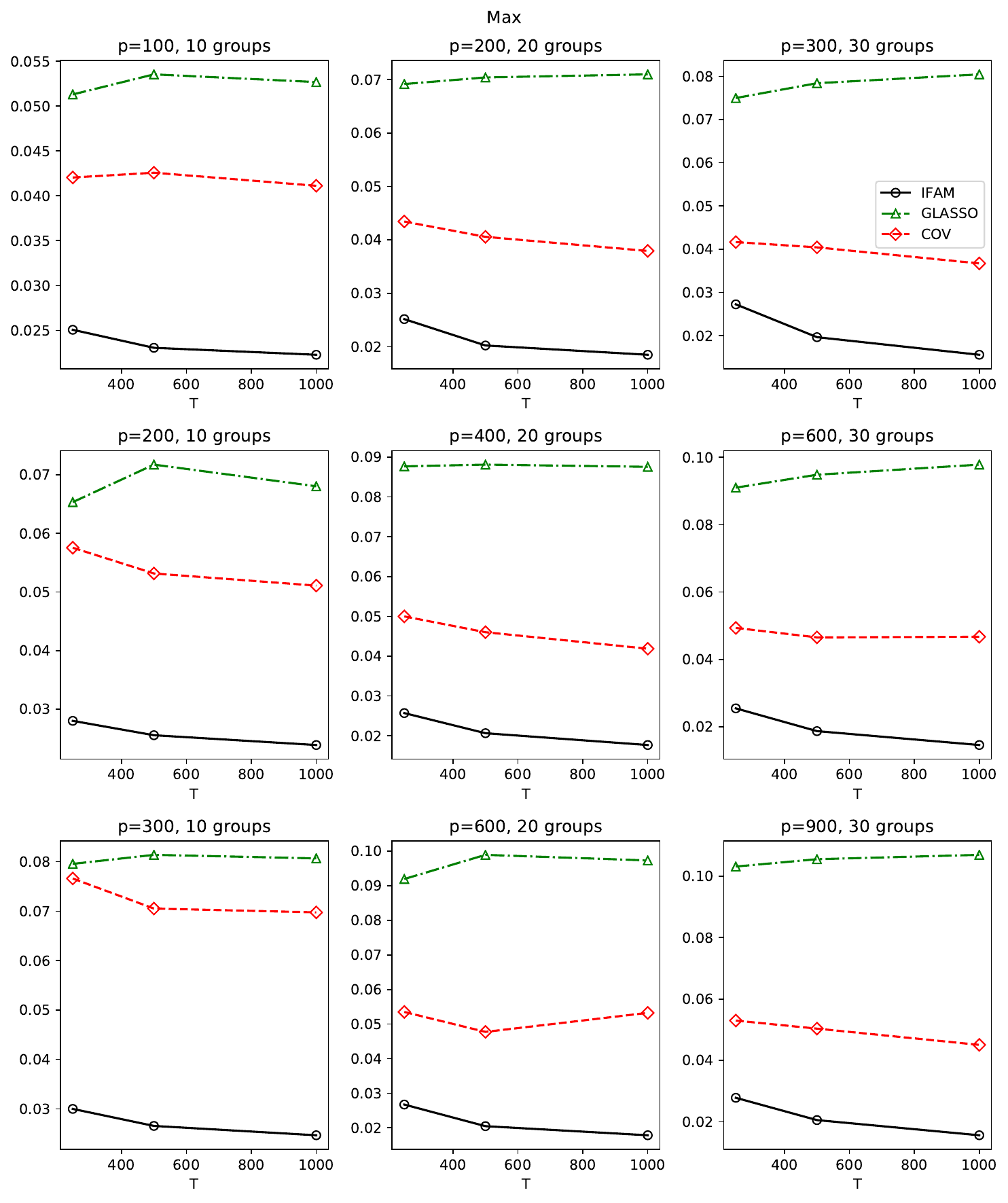}
\caption{Mean matrix max norm for the large matrix estimation error for various $p$ and the number of groups.} \label{fig:whole-max}
\end{figure}

\begin{figure}[!h]
\centering
\includegraphics[width = 0.8\textwidth]{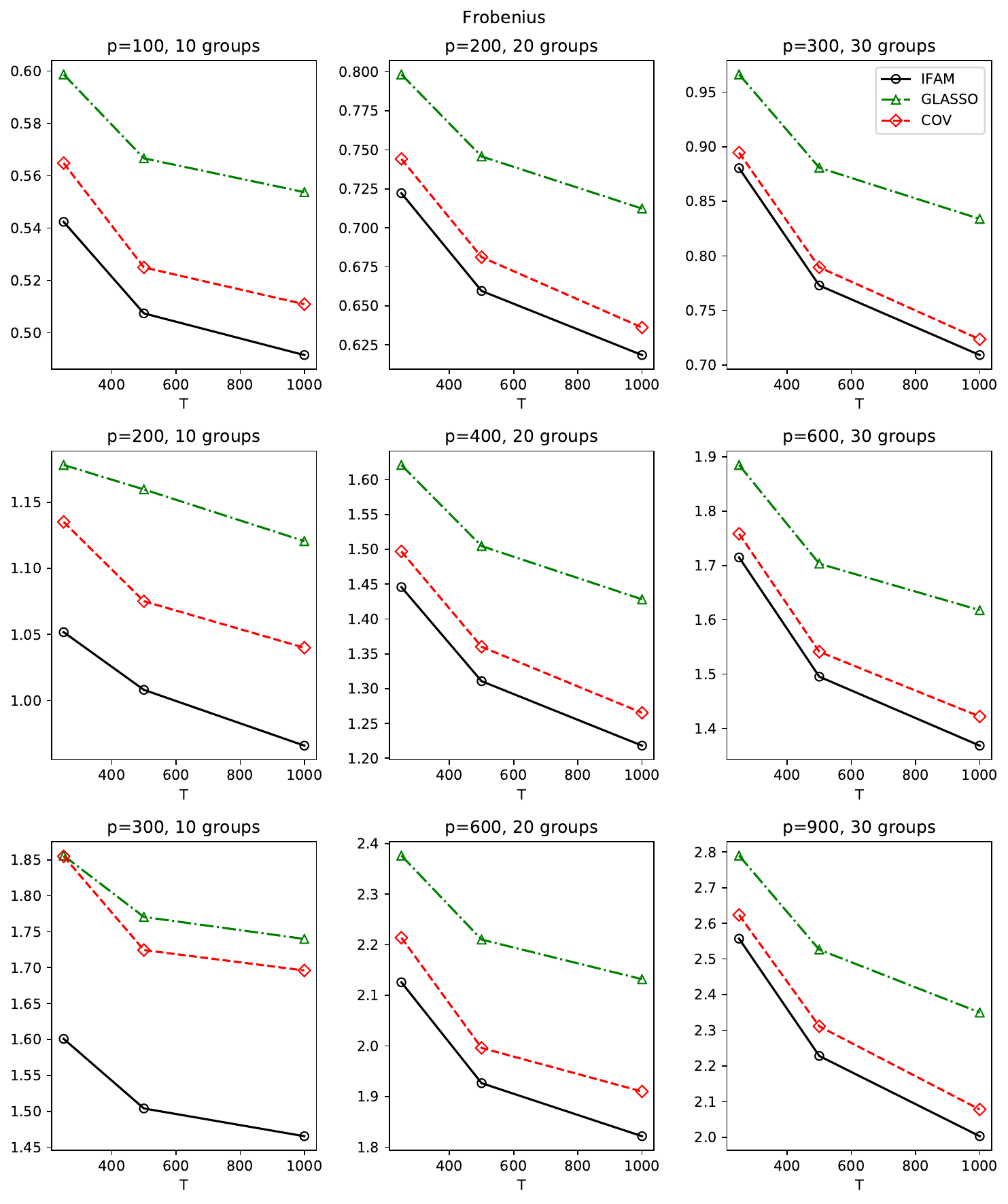}
\caption{Mean matrix Frobenius norm for the large matrix estimation error for various $p$ and the number of groups.} \label{fig:whole-fro}
\end{figure}

\begin{figure}[!h]
\centering
\includegraphics[width = 0.8\textwidth]{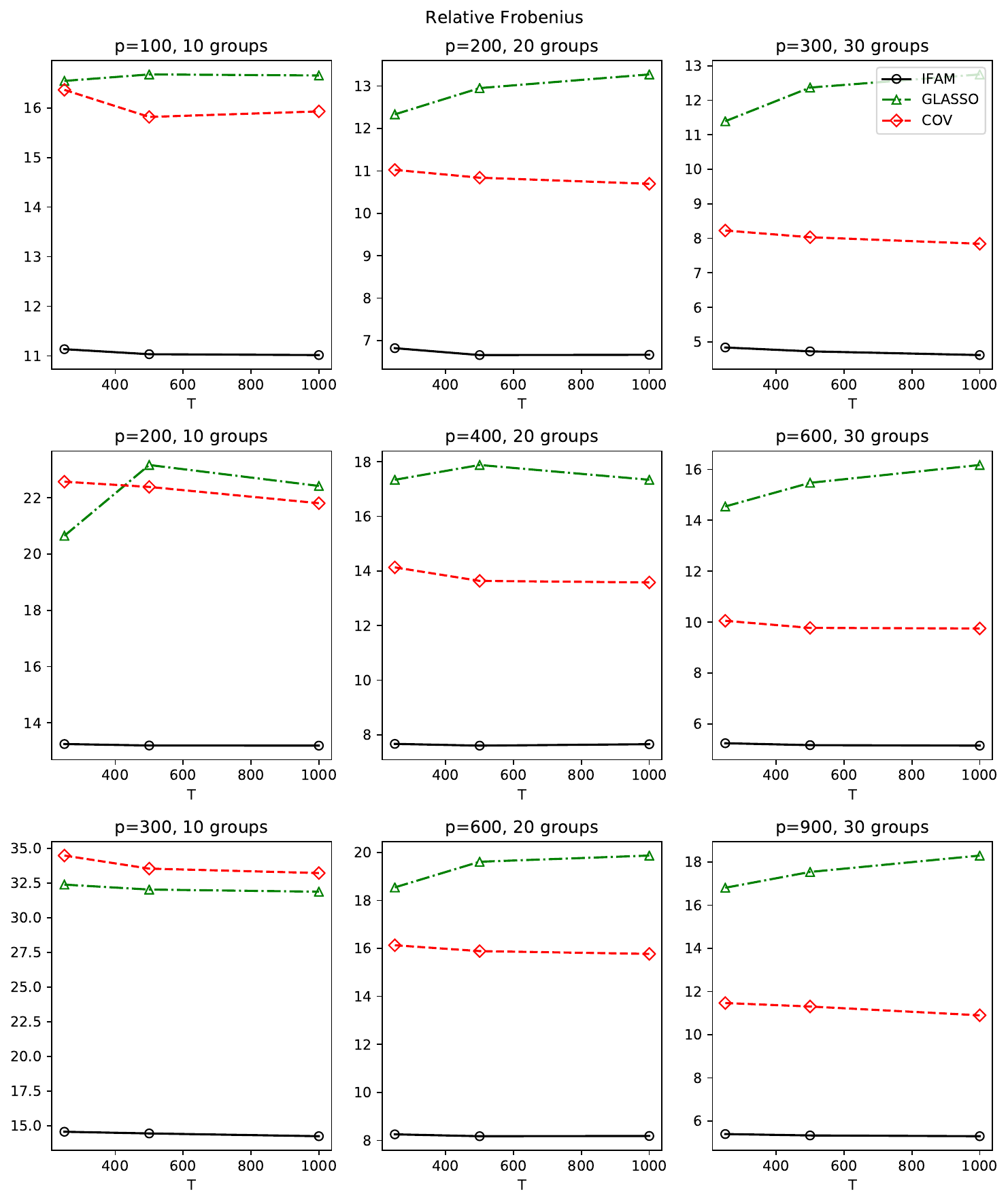}
\caption{Mean matrix relative Frobenius norm for the large matrix estimation error for various $p$ and the number of groups.} \label{fig:whole-rel}
\end{figure}

Figures \ref{fig:whole-max}, \ref{fig:whole-fro}, and \ref{fig:whole-rel} draw the mean matrix max, Frobenius, and relative Frobenius norms for the matrix estimation errors, respectively.
From Figures \ref{fig:whole-max}, \ref{fig:whole-fro}, and \ref{fig:whole-rel}, we find that the matrix errors decrease as the sample size increases, and IFAM outperforms the other benchmarks.
These results are consistent with the main simulation analysis.

\begin{figure}[!h]
\centering
\includegraphics[width = 0.8\textwidth]{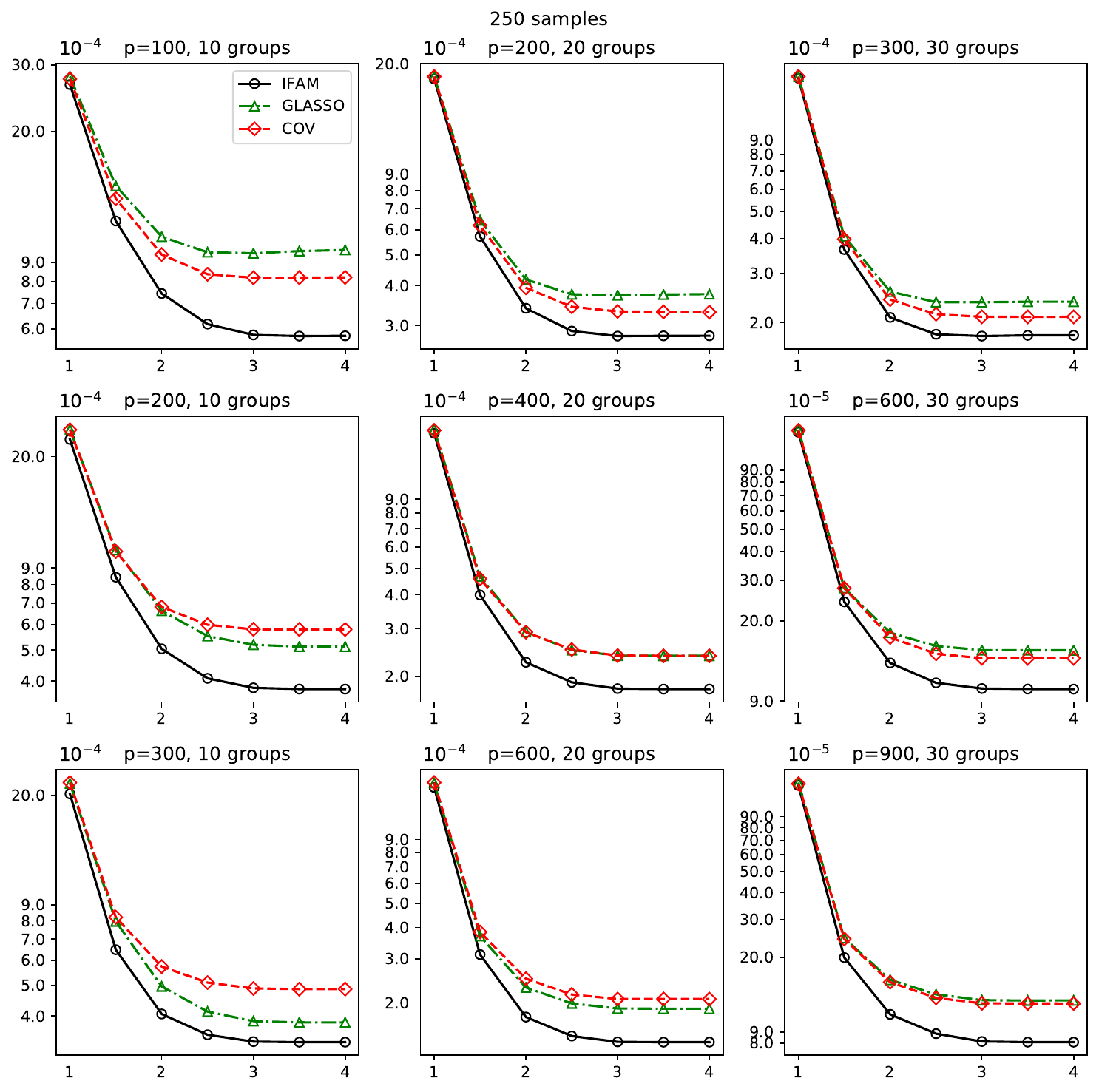}
\caption{Average expected out-of-sample portfolio risk using the Double-POET estimation with group memberships obtained from IFAM, GLASSO, and COV, against the gross exposure constraint $c_0$ for varying the number of clusters and cluster size from 10 to 30 for $T=250$.}
\label{fig:risk250}
\end{figure}

\begin{figure}[!h]
\centering
\includegraphics[width = 0.8\textwidth]{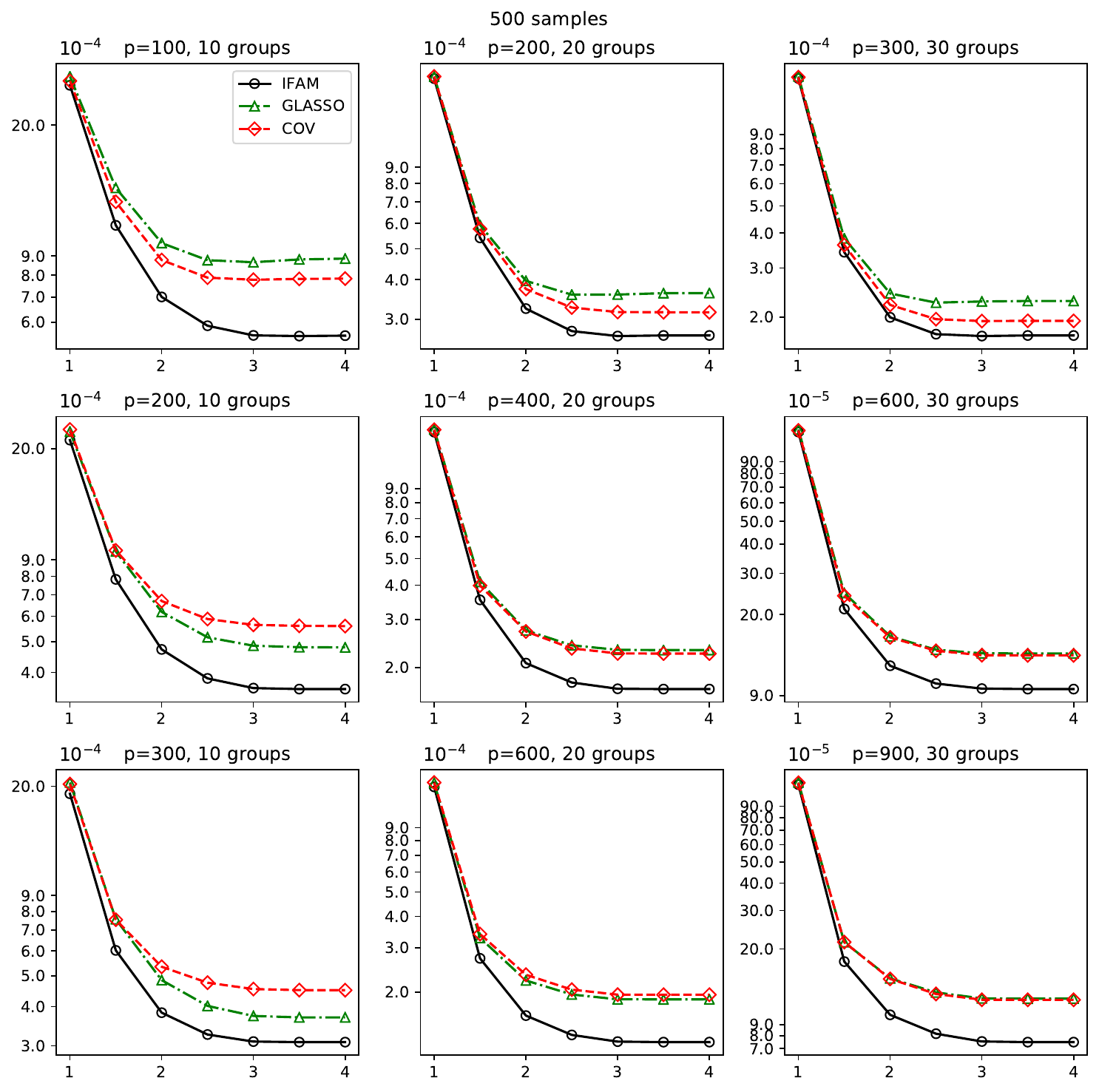}
\caption{Average expected out-of-sample portfolio risk using the Double-POET estimation with group memberships obtained from IFAM, GLASSO, and COV against the gross exposure constraint $c_0$ for varying the number of clusters and cluster size from 10 to 30 for $T=500$.}
\label{fig:risk500}
\end{figure}

\begin{figure}[!h]
\centering
\includegraphics[width = 0.8\textwidth]{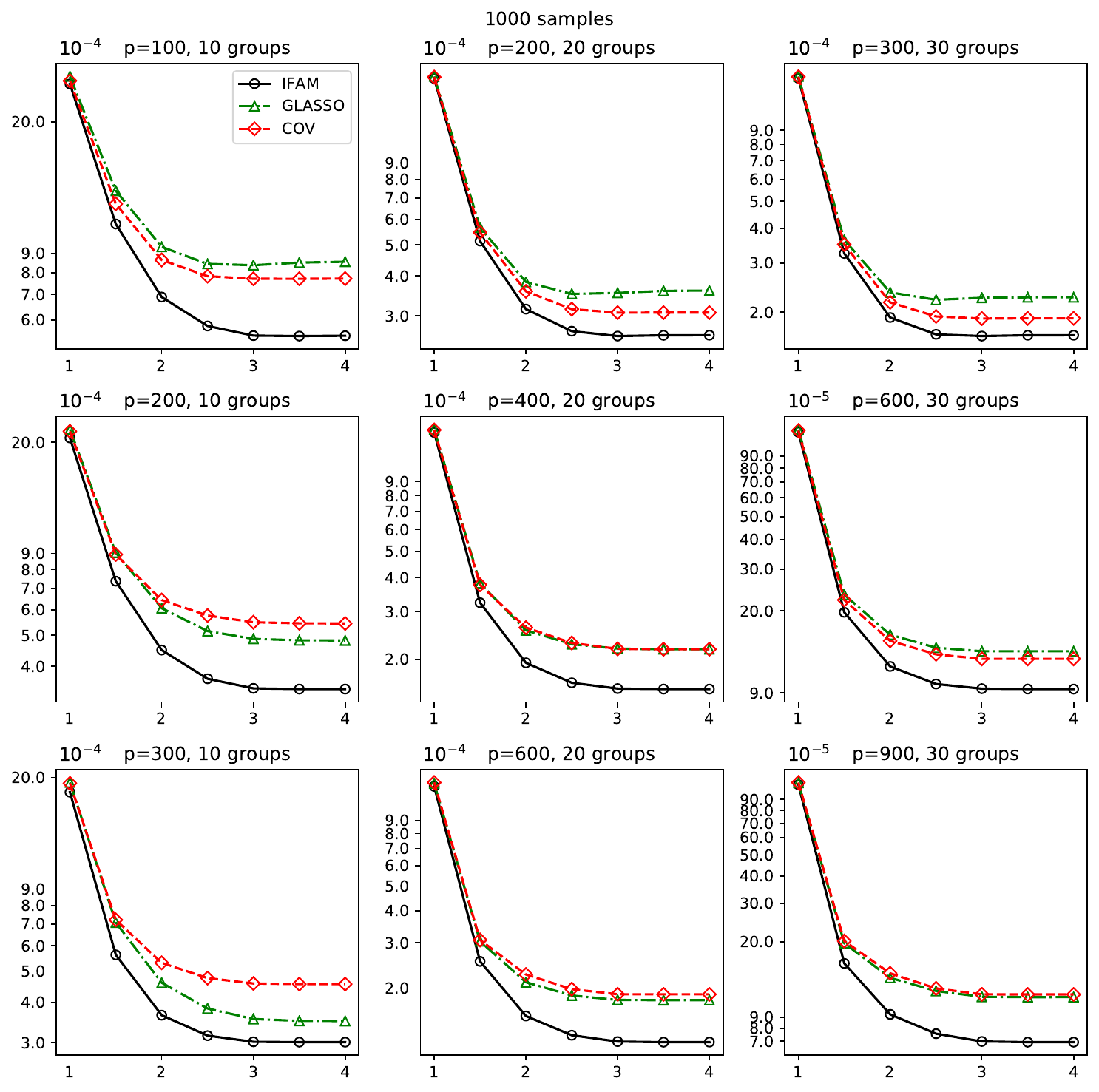}
\caption{Average expected out-of-sample portfolio risk using the Double-POET estimation with group memberships obtained from IFAM, GLASSO, and COV against the gross exposure constraint $c_0$ for varying the number of clusters and cluster size from 10 to 30 for $T=1000$.}
\label{fig:risk1000}
\end{figure}

Figures \ref{fig:risk250}, \ref{fig:risk500}, and \ref{fig:risk1000} draw the average expected out-of-sample risk using the Double-POET estimation with group memberships obtained from IFAM, GLASSO, and COV, against the gross exposure constraint $c_0$ for varying the number of clusters and cluster size from 10 to 30 for $T=250,500,$ and $1000$, respectively.
From Figures \ref{fig:risk250}, \ref{fig:risk500}, and \ref{fig:risk1000}, we find that IFAM achieves the lowest average expected out-of-sample risk across all specifications.
From these results, we can conclude that IFAM provides better group membership identification compared to other benchmarks, regardless of the number of clusters or cluster size, which leads to improved matrix estimation and portfolio allocation performance.

\clearpage

\section{Clustering algorithm}\label{sec:clustering}
For the group membership identification, we use the regularized spectral clustering algorithm \citep{qin2013regularized} as in Algorithm \ref{rsc}.
\begin{algorithm}
  \caption{Regularized spectral clustering algorithm}   \label{rsc}
  \begin{algorithmic}
  \State \textbf{Step 1 } Estimate the regularized graph Laplacian:
  \begin{equation*}
    L_{\tau} = D_{\tau}^{-1/2} A D_{\tau}^{-1/2}
    ,
  \end{equation*}
  where $A$ is a given weighted adjacency matrix, $D$ is a diagonal degree matrix with $D_{ii} = \sum_{j} | [A]_{ij}|$, $\tau = \frac{1}{p} \sum_{i=1}^{p}  D_{ii}$ is a regularization parameter, $D_{\tau}$ is a regularized diagonal degree matrix with $[D_{\tau}]_{i,i}=D_{ii}+\tau$.
  
  \State \textbf{Step 2 } Find the eigenvectors $U_1, \ldots, U_{K} \in \mathbb{R}^{p}$ corresponding to the $K$ largest eigenvalues of $L_{\tau}$

  \State \textbf{Step 3 } Estimate the normalized eigenvectors $\tilde{U} \in \mathbb{R}^{p \times (K-1)}$ by normalizing each of $U$'s rows to have unit length, where $U = [U_{2},\ldots, U_{K}] \in \mathbb{R}^{p \times (K-1)}$ ($\tilde{U}_{ij}=U_{ij}/(\sum_{j} U_{ij}^2)^{1/2}$).

  \State \textbf{Step 4 } Conduct k-means clustering on the $K$-dimensional feature matrix, $\tilde{U}$, using $K$ clusters.

  \end{algorithmic}
\end{algorithm}

\section{Proofs}\label{sec:proofs}

\subsection{Proof of Theorem \ref{prop:density}}

\begin{lemma}\label{lemma:main-term}
  Under Assumption \ref{ass:sparsity}, for any $G \in \mathcal{G}$, there exists a set $A \subset G$ such that $\abs{A} \geq \frac{\gamma_{G} \abs{G}}{s_g^2 (r_c + r_G)^3}$ and each element in $\mathcal{K}_{G,A}$ is greater than $\delta_{G} |G|^{-1}$,
  where
  \begin{eqnarray*}
    && \delta_G = \frac{\sigma_{\min}^{2}([\bSigma_{u}^{-1}]_{G})  \sqrt{r_c + r_G} }{4  \sigma_{\max}(\bB^{\top} \bSigma_{u}^{-1} \bB)} , \quad
    \gamma_G = \frac{  \sigma^{2}_{\min}([\bSigma_{u}^{-1}]_{G}) }{16  \norm{[\bSigma_{u}^{-1}]_{G}}_{\max}^{2} \sigma_{\max}(\bB^{\top} \bSigma_{u}^{-1} \bB) C_G} , \cr 
    && C_G = \norm{[\bB]_{G,\cdot}}_{\max}^{2} |G| , \quad
    \mathcal{K}_{G} =  [\bSigma_{u}^{-1}]_{G} [\bB]_{G,\cdot} \mathcal{Y} [\bB]_{G,\cdot}^{\top} [\bSigma_{u}^{-1}]_{G} , \cr
    && \mathcal{Y} = (\bB^{\top} \bSigma_u^{-1}  \bB)^{-1} - (\bB^{\top} \bSigma_u^{-1}  \bB)^{-1} {\bSigma_{F}}^{-1} (\bB^{\top} \bSigma_u^{-1}  \bB)^{-1} , \text{ and } \mathcal{K}_{G,A} = \left[ \mathcal{K}_{G} \right]_{A}
    .
  \end{eqnarray*}
\end{lemma}

\textbf{Proof of Lemma \ref{lemma:main-term}}.
Let $\mathcal{J}$ be the set of non-zero vector columns of $[\bB]_{G,\cdot}$.
Simple algebra shows that
\begin{equation*}
  \mathcal{K}_{G} =  [\bSigma_{u}^{-1}]_{G} [\bB]_{G,\mathcal{J}} [\mathcal{Y}]_{\mathcal{J}} [\bB]_{G,\mathcal{J}}^{\top} [\bSigma_{u}^{-1}]_{G} 
  .
\end{equation*}
Assumptions \ref{ass:sparsity}\ref{sparsity:nontrivial-contribution}--\ref{sparsity:sparsity} imply that there exist $C_1$ and $C_2$ such that $C_1 < \sigma_{\min}(\bB^{\top}\bSigma_{u}^{-1}\bB) \leq \sigma_{\max}(\bB^{\top}\bSigma_{u}^{-1}\bB) < C_2$.
By Assumption \ref{ass:sparsity}\ref{sparsity:factor-size}, there exists $C_3$ such that $\sigma_{\max}(\bSigma_{F}^{-1}) < C_3 p^{-v}$.
Therefore, by Weyl's theorem, we have
\begin{eqnarray}\label{eq:Y-spectral}
  \frac{1}{2} \sigma_{\max}^{-1}(\bB^{\top} \bSigma_u^{-1}  \bB) < \sigma_{\min}(\mathcal{Y}) \leq  \sigma_{\max}(\mathcal{Y}) < 2 \sigma_{\min}^{-1}(\bB^{\top} \bSigma_u^{-1}  \bB)
\end{eqnarray}
for sufficiently large $p$.
Let $\mathcal{V}_{\mathcal{J}} \bSigma_{\mathcal{V}} \mathcal{V}_{\mathcal{J}}^{\top}$ be the eigendecomposition of $[\mathcal{Y}]_{\mathcal{J}}$.
Using the fact that $\tr (M_2^{\top} M_1 M_2) $ $\geq \sigma_{\min} (M_1) \tr (M_2^{\top} M_2)$ for any positive definite matrix $M_1$ and conformable matrix $M_2$,
we have
\begin{eqnarray}\label{eq:min-size}
  \norm{[\bSigma_u^{-1}]_{G} [\bB]_{G, \mathcal{J}} \mathcal{V}_{\mathcal{J}} \bSigma_{\mathcal{V}}^{1/2}}_{F}^{2}
  &=& 
  \tr (\bSigma_{\mathcal{V}}^{1/2} \mathcal{V}_{\mathcal{J}}^{\top} [\bB]_{G, \mathcal{J}}^{\top} ([\bSigma_u^{-1}]_{G})^{2} [\bB]_{G, \mathcal{J}} \mathcal{V}_{\mathcal{J}} \bSigma_{\mathcal{V}}^{1/2} ) \cr
  &\geq& \sigma_{\min}^{2}([\bSigma_{u}^{-1}]_{G}) \tr ( \bSigma_{\mathcal{V}}^{1/2} \mathcal{V}_{\mathcal{J}}^{\top} [\bB]_{G, \mathcal{J}}^{\top}  [\bB]_{G, \mathcal{J}} \mathcal{V}_{\mathcal{J}} \bSigma_{\mathcal{V}}^{1/2} ) \cr
  &=& \sigma_{\min}^{2}([\bSigma_{u}^{-1}]_{G}) \tr (   [\bB]_{G, \mathcal{J}} [\mathcal{Y}]_{\mathcal{J}} [\bB]_{G, \mathcal{J}}^{\top}) \cr
  &\geq& \frac{1}{2} \sigma_{\min}^{2}([\bSigma_{u}^{-1}]_{G}) \sigma_{\max}^{-1}(\bB^{\top} \bSigma_{u}^{-1} \bB) \norm{[\bB]_{G,\mathcal{J}}}_{F}^{2} \cr
  &\geq& \frac{1}{2} \sigma_{\min}^{2}([\bSigma_{u}^{-1}]_{G}) \sigma_{\max}^{-1}(\bB^{\top} \bSigma_{u}^{-1} \bB) |\mathcal{J}|
  ,
\end{eqnarray}
where the second and last equalities are by \eqref{eq:Y-spectral} and Assumption \ref{ass:sparsity}\ref{sparsity:normalized-pervasive}, respectively.
On the other hand, for the $i$th row of $[\bSigma_u^{-1}]_{G} [\bB]_{G, \mathcal{J}} \mathcal{V}_{\mathcal{J}} \bSigma_{\mathcal{V}}^{1/2}$, we have
\begin{eqnarray}\label{eq:emax-multiple}
  \norm{[[\bSigma_u^{-1}]_{G} [\bB]_{G, \mathcal{J}} \mathcal{V}_{\mathcal{J}}  \bSigma_{\mathcal{V}}^{1/2}]_{i,\cdot} }_{2}^{2} &=&  \sum_{j \in \mathcal{J}}  \left(  \sum_{l \in \mathcal{J}} \sum_{k\in G} [[\bSigma_{u}^{-1}]_{G}]_{i,k} \bB_{k,l} [\mathcal{V}_{\mathcal{J}} \bSigma_{\mathcal{V}}^{1/2}]_{l,j}   \right)^2   \cr
  &\leq& |\mathcal{J}|  \sum_{j \in \mathcal{J}}    \sum_{l \in \mathcal{J}} \left( \sum_{k\in G} [[\bSigma_{u}^{-1}]_{G}]_{i,k} \bB_{k,l} [\mathcal{V}_{\mathcal{J}} \bSigma_{\mathcal{V}}^{1/2}]_{l,j}   \right)^2   \cr
  &\leq& s_g |\mathcal{J}|  \sum_{j \in \mathcal{J}}    \sum_{l \in \mathcal{J}}  \sum_{\substack{k\in G \text{ s.t.} \cr [\bSigma_{u}^{-1}]_{i,k} \neq 0}} \left( [[\bSigma_{u}^{-1}]_{G}]_{i,k} \bB_{k,l} [\mathcal{V}_{\mathcal{J}} \bSigma_{\mathcal{V}}^{1/2}]_{l,j}   \right)^2   \cr
  &\leq& s_g^2 |\mathcal{J}|^{3} C_{\mathcal{J}}^2 \norm{[\bSigma_{u}^{-1}]_{G}}_{\max}^{2}  \norm{\mathcal{V}_{\mathcal{J}} \bSigma_{\mathcal{V}}^{1/2}}_{\max}^{2} \cr
  &\leq& 2 s_g^2 |\mathcal{J}|^{3} C_{\mathcal{J}}^2 \norm{[\bSigma_{u}^{-1}]_{G}}_{\max}^{2}  \sigma_{\min}^{-1} (\bB^{\top} \bSigma_{u}^{-1} \bB)
  ,
\end{eqnarray}
where $C_{\mathcal{J}} =  \norm{[\bB]_{G,\mathcal{J}}}_{\max} = O(p^{-v/2})$ and the first and second inequalities follow from Jensen's inequality and Assumption \ref{ass:sparsity}\ref{sparsity:sparsity}.
Let $\delta = \frac{\sigma_{\min}^{2}([\bSigma_{u}^{-1}]_{G})  |\mathcal{J}| }{4|G|  \sigma_{\max}(\bB^{\top} \bSigma_{u}^{-1} \bB)}$.
We have
\begin{align}\label{eq:max-size}
  &\norm{[\bSigma_u^{-1}]_{G} [\bB]_{G, \mathcal{J}} \mathcal{V}_{\mathcal{J}} \bSigma_{\mathcal{V}}^{1/2} }_{F}^{2} \cr
  &= 
  \sum_{i \in \mathcal{A}} \norm{[[\bSigma_u^{-1}]_{G} [\bB]_{G, \mathcal{J}} \mathcal{V}_{\mathcal{J}} \bSigma_{\mathcal{V}}^{1/2} ]_{i,\cdot}}_{2}^{2}   + \sum_{i \in G \setminus \mathcal{A}}  \norm{[[\bSigma_u^{-1}]_{G} [\bB]_{G, \mathcal{J}} \mathcal{V}_{\mathcal{J}} \bSigma_{\mathcal{V}}^{1/2} ]_{i,\cdot}}_{2}^{2} \cr
  &\leq  \sum_{i \in \mathcal{A}}  \norm{[[\bSigma_u^{-1}]_{G} [\bB]_{G, \mathcal{J}} \mathcal{V}_{\mathcal{J}} \bSigma_{\mathcal{V}}^{1/2} ]_{i,\cdot}}_2^2 + | G \setminus \mathcal{A} |   \delta \cr
  &\leq \sum_{i \in \mathcal{A}} \norm{[[\bSigma_u^{-1}]_{G} [\bB]_{G, \mathcal{J}} \mathcal{V}_{\mathcal{J}} \bSigma_{\mathcal{V}}^{1/2} ]_{i,\cdot}}_2^2 + \frac{1}{4}  \sigma_{\min}^{2}([\bSigma_{u}^{-1}]_{G}) \sigma_{\max}^{-1}(\bB^{\top} \bSigma_{u}^{-1} \bB) |\mathcal{J}| 
  ,
\end{align}
where $\mathcal{A} = \left\lbrace i \in G \Big| \norm{[[\bSigma_u^{-1}]_{G} [\bB]_{G, \mathcal{J}} \mathcal{V}_{\mathcal{J}} \bSigma_{\mathcal{V}}^{1/2}  ]_{i,\cdot}}_{2}^{2} > \delta  \right\rbrace$.
Using \eqref{eq:min-size} and \eqref{eq:max-size}, we have
\begin{eqnarray*}
  \frac{1}{4}  \sigma_{\min}^{2}([\bSigma_{u}^{-1}]_{G}) \sigma_{\max}^{-1}(\bB^{\top} \bSigma_{u}^{-1} \bB) |\mathcal{J}|  &\leq&  \sum_{i \in \mathcal{A}} \norm{[[\bSigma_u^{-1}]_{G} [\bB]_{G, \mathcal{J}} \mathcal{V}_{\mathcal{J}}  \bSigma_{\mathcal{V}}^{1/2}]_{i,\cdot}}_2^2 \cr
  &\leq& 2 s_g^2 |\mathcal{J}|^3 C_{\mathcal{J}}^{2} \norm{[\bSigma_{u}^{-1}]_{G}}_{\max}^2  \sigma_{\min}^{-1} (\bB^{\top} \bSigma_{u}^{-1} \bB) |\mathcal{A}| 
  ,
\end{eqnarray*}
where the second inequality is due to \eqref{eq:emax-multiple}.
Therefore, we have
\begin{equation}\label{eq:mathA-prop}
  |\mathcal{A}| \geq \frac{  \sigma^{2}_{\text{min}}([\bSigma_{u}^{-1}]_{G}) }{8s_g^2 |\mathcal{J}|^{2} \norm{[\bSigma_{u}^{-1}]_{G}}_{\max}^{2}C_\mathcal{J}^2 \sigma_{\max}(\bB^{\top} \bSigma_{u}^{-1} \bB)}  =  \frac{2 \gamma_{G} |G|}{s_g^{2} (r_c + r_G)^{2}}
  .
\end{equation}

Using the fact that $\mathcal{K}_{G} = ([\bSigma_u^{-1}]_{G} [\bB]_{G, \mathcal{J}} \mathcal{V}_{\mathcal{J}} \bSigma_{\mathcal{V}}^{1/2}) ([\bSigma_u^{-1}]_{G} [\bB]_{G, \mathcal{J}} \mathcal{V}_{\mathcal{J}} \bSigma_{\mathcal{V}}^{1/2})^{\top}$, we have
\begin{equation*}
  [\mathcal{K}_{G,\mathcal{A}}]_{i,j} = \psi_{i}^{\top} \psi_{j}
  ,
\end{equation*}
where $\psi_{i}, \psi_{j} \in \Psi_{G,\mathcal{A}} = \lbrace [[\bSigma_u^{-1}]_{G} [\bB]_{G, \mathcal{J}} \mathcal{V}_{\mathcal{J}} \bSigma_{\mathcal{V}}^{1/2}]_{i,\cdot} \in \mathbb{R}^{|\mathcal{J}|} | i \in \mathcal{A} \rbrace$.
We can consider a set of points $\left\lbrace \psi/\left\lVert \psi \right\rVert _2 | \psi \in \Psi_{G,\mathcal{A}} \right\rbrace$ on the unit sphere in $\mathbb{R}^{|\mathcal{J}|}$.
By Theorem 1.1 in \citet{boroczky2003covering} and the pigeonhole principle, there exists a set $A \subset \mathcal{A}$ such that $\abs{A} \geq \frac{\abs{\mathcal{A}}}{2|\mathcal{J}|}$  and $(\psi_i^{\top} \psi_j) / (\norm{\psi_i}_{2} \norm{\psi_j}_{2}) \geq |\mathcal{J}|^{-1/2}$ for any $i, j \in A$.
Therefore, for any $i,j \in A$, we have
\begin{eqnarray*}
  [\mathcal{K}_{G,A}]_{i,j} &\geq&  \psi_{i}^{\top} \psi_{j} \cr
  &\geq& \frac{\psi_{i}^{\top} \psi_{j}}{\norm{\psi_i}_{2} \norm{\psi_j}_{2}} \norm{\psi_i}_{2} \norm{\psi_j}_{2} \cr
  &\geq& |\mathcal{J}|^{-1/2}  \delta
  .
\end{eqnarray*}
This fact and \eqref{eq:mathA-prop} conclude that there exists a set $A \subset G$ such that $\abs{A} \geq \frac{\gamma_{G} \abs{G}}{s_g^2 (r_c + r_G)^3}$ and each element in $\mathcal{K}_{G,A}$ is greater than $\delta_{G} |G|^{-1}$.
$\blacksquare$

\begin{lemma}\label{lemma:chi-term}
  Under Assumption \ref{ass:sparsity}, we have
  \begin{eqnarray*}
    [\chi]_{i,j}=
    \begin{cases}
      O(p^{-(1-v)/2} + s_b p^{-v}) , & \text{if } (i,j) \in \mathcal{O} ,\cr
      O(1) , & \text{otherwise,}
    \end{cases}
  \end{eqnarray*}
  where $\mathcal{O} = \left\lbrace (i,j) |  r_c + 1 \leq i,j \leq r \text{ and }  h(i) \neq h(j) \right\rbrace$, $h$ is defined in \eqref{eq:concatenated-loading}, and $\chi = (\bB^{\top} \bSigma_u \bB)^{-1}$.
\end{lemma}

\textbf{Proof of Lemma \ref{lemma:chi-term}.}
Since $C_1 <\sigma_{\min} (\bB^{\top} \bSigma_u^{-1} \bB) \leq \sigma_{\max} (\bB^{\top} \bSigma_u^{-1} \bB) < C_2$ for some positive constants $C_1$ and $C_2$, we have $\left \lVert \chi \right \rVert _{\max} \leq C_1^{-1}$.
Specifically, we have
\begin{equation}\label{eq:chi-same-factor}
  |[\chi]_{i,j} | \leq C_{1}^{-1} \text{ for all } (i,j) \in \left\lbrace 1, \ldots, r \right\rbrace^2 \setminus \mathcal{O}.
\end{equation}
Let $\mathcal{D} = \lbrace (i,j) | 1 \leq i,j \leq r_c  \rbrace \cup \lbrace (i,j) | \bar{r}_{k-1} < i,j \leq \bar{r}_{k} \text{ for some } 1 \leq k \leq J \rbrace$ and $\Phi = \bB^{\top} \bSigma_u^{-1} \bB = \Phi_{d} + \Phi_{e}$, where $\Phi_{d}$ is the block-diagonal matrix of $\Phi$, that is, $[\Phi_d]_{i,j}=\Phi_{i,j}$ for $(i,j) \in \mathcal{D}$ and $[\Phi_d]_{i,j}=0$ for $(i,j) \notin \mathcal{D}$, and $\Phi_{e}$ is the corresponding off-block-diagonal matrix of $\Phi$.
Then, we have $C_1 < \sigma_{i} ([\Phi_d]) < C_2$ for all $1 \leq i \leq r$.
Consider $[\Phi_e]_{i,j}$ for any $i,j$, and $k$ such that $1 \leq i \leq r_c$, $\bar{r}_{G_{k-1}} < j \leq \bar{r}_{G_{k}}$, and $1 \leq k \leq J$.
There exists a positive constant $C_3$ such that
\begin{eqnarray}\label{eq:Phi-gl}
  \left|[\Phi_e]_{i,j}\right|  &=&  \left|\sum_{x=1}^{p} \sum_{y \in G_k} [\bB]_{x,i} [\bSigma_u^{-1}]_{x,y} [\bB]_{y,j}\right|  \cr
  &=& \left|\sum_{x \in G_k} \sum_{y \in G_k} [\bB]_{x,i} [\bSigma_u^{-1}]_{x,y} [\bB]_{y,j} + \sum_{x \in G_k^{c}} \sum_{y \in G_k} [\bB]_{x,i} [\bSigma_u^{-1}]_{x,y} [\bB]_{y,j}\right|  \cr
  &\leq& C(s_g p^{-(1-v)/2}) + C(p^{-(1+v)/2} s_b) \cr
  &\leq& C_3  p^{-(1-v)/2}
  ,
\end{eqnarray}
where the first inequality is by Assumption \ref{ass:sparsity}\ref{sparsity:sparsity}.
Consider $r_c < i,j \leq r$ and $h(i) \neq h(j)$. %
We have
\begin{eqnarray}\label{eq:Phi-ll}
  \left|[\Phi_e]_{i,j}\right|  &=&  \left|\sum_{x \in G_{h(i)}} \sum_{y \in G_{h(j)}} [\bB]_{x,i} [\bSigma_u^{-1}]_{x,y} [\bB]_{y,j}\right|  \cr
  &=& O(s_b p^{-v})
  ,
\end{eqnarray}
where the second equality is by Assumption \ref{ass:sparsity}\ref{sparsity:sparsity}.
Furthermore, using Assumption \ref{ass:sparsity}\ref{sparsity:sparsity}, we can show
\begin{equation}\label{eq:Phi-l-sparse}
  \max_{r_c+1 \leq i \leq r} \sum_{j=1}^{r} \b1([\Phi_e]_{i,j} \neq 0) \leq r_c + \tilde{r} + \tilde{r} s_b \leq 3 \tilde{r} s_b
  ,
\end{equation}
where $\tilde{r} = \max (r_c, r_1, r_2, \ldots, r_J)$.
Using the fact that $\Phi = (I + \Phi_e \Phi_d^{-1}) \Phi_d$, we have
\begin{eqnarray*}
  \Phi^{-1} &=&  \Phi_d^{-1} (I+\Phi_e \Phi_d^{-1})^{-1} \cr
  &=& \Phi_d^{-1} (I - \Phi_e \Phi_d^{-1} + \Phi_e \Phi_d^{-1}\Phi_e \Phi_d^{-1} - \cdots) \cr
  &=& \Phi_d^{-1}  -  \Phi_d^{-1} \Phi_e \Phi_d^{-1} + \Phi_d^{-1} \Phi_e \Phi_d^{-1}\Phi_e \Phi_d^{-1} - \cdots \cr
  &=& \sum_{k=0}^{\infty} \tilde{\Phi}_{k}
  ,
\end{eqnarray*}
where $\tilde{\Phi}_{0} = \Phi_{d}^{-1}$, $\tilde{\Phi}_{k} = \tilde{\Phi}_{k-1} P$ for $k\geq 1$, $P=-\Phi_e \Phi_d^{-1}$, and the second equality follows from recursively applying the Sherman-Morrison-Woodbury matrix identity.
Since $\Phi_{d}^{-1}$ is a block-diagonal matrix, $P$ satisfies the properties of $\Phi_{e}$, i.e. \eqref{eq:Phi-gl}, \eqref{eq:Phi-ll}, and \eqref{eq:Phi-l-sparse}.
Let
\begin{equation*}
  \tilde{\Phi}_{k} =
  \begin{pmatrix}
    \mathcal{W}_{k,1} & \mathcal{W}_{k,3}^{\top} \cr
    \mathcal{W}_{k,3} & \mathcal{W}_{k,2}
  \end{pmatrix}
  \text{ and }
  P =
  \begin{pmatrix}
    P_{1} & P_{3}^{\top} \cr
    P_{3} & P_{2}
  \end{pmatrix}
  ,
\end{equation*}
where $\mathcal{W}_{k,1}, P_1 \in \mathbb{R}^{r_c \times r_c}$, $\mathcal{W}_{k,2}, P_2 \in \mathbb{R}^{(r-r_c) \times (r-r_c)}$, and $\mathcal{W}_{k,3}, P_3 \in \mathbb{R}^{(r-r_c)\times r_c}$.
Simple algebra shows that $P_1=0$.
Then, we have
\begin{eqnarray*}
  && \mathcal{W}_{k+1,2} = \mathcal{W}_{k,3} P_{3}^{\top} + \mathcal{W}_{k,2} P_2 .
\end{eqnarray*}
We have
\begin{eqnarray*}
  && \sum_{k=0}^{\infty}  \mathcal{W}_{k+1,2} = \left( \sum_{k=0}^{\infty} \mathcal{W}_{k,3} \right)  P_{3}^{\top} + \left( \sum_{k=0}^{\infty} \mathcal{W}_{k,2}  \right) P_2   . %
\end{eqnarray*}
Then, for $r_c <  i,j \leq r$ and $h(i) \neq h(j)$, we have
\begin{eqnarray*}
  \left|[\chi]_{i,j}\right|  &=& \left|[\Phi_d^{-1}]_{i,j} + \sum_{k=0}^{\infty}  [\mathcal{W}_{k+1,2}] _{i-r_c, j-r_c}\right|  \cr
  &=& \left|\left(  \sum_{k=0}^{\infty} [\mathcal{W}_{k,3}]_{i - r_c,\cdot} \right)    [P_{3}^{\top}]_{\cdot,j-r_c}\right|  +  \left|\left( \sum_{k=0}^{\infty} [\mathcal{W}_{k,2}]_{i-r_c,\cdot}  \right) [P_2]_{\cdot,j-r_c} \right|   \cr
  &\leq& C p^{-(1-v)/2} + C s_b p^{-v}
  ,
\end{eqnarray*}
where the last inequality follows from \eqref{eq:chi-same-factor}, \eqref{eq:Phi-gl}, \eqref{eq:Phi-ll}, and \eqref{eq:Phi-l-sparse}.
Therefore, we have
\begin{eqnarray*}
  [\chi]_{i,j}=
  \begin{cases}
    O(p^{-(1-v)/2} + s_b p^{-v}) , & \text{if } r_c + 1 \leq i,j \leq r \text{ and }  h(i) \neq h(j) ,\cr
    O(1) , & \text{otherwise.}
  \end{cases}
\end{eqnarray*}
$\blacksquare$

\textbf{Proof of Theorem \ref{prop:density}.}
It is sufficient to show that
\begin{eqnarray}
  && \left|G \times G \right|^{-1}  \sum_{(i,j) \in G \times G} \b1 ([{\bSigma^{-1}}]_{i,j} \leq - \delta_G |G|^{-1} + \epsilon_{1,p} ) > \frac{\gamma_{G}^{2}}{s_g^4(r_c+r_G)^6} - \epsilon_{2,p} \quad \label{eq:thm1-original-version-within} \\
  &\text{and}& \left|G \times G^c \right|  ^{-1} \sum_{(i,j) \in G \times G^{c}} \b1 (|[{\bSigma^{-1}}]_{i,j}| \geq \epsilon_{3,p}) < \epsilon_{4,p} \label{eq:thm1-original-version-between}
  ,
\end{eqnarray}
where $\epsilon_{1,p}=o(p^{-v})$, $\epsilon_{2,p}=O(s_g p^{-v})$, $\epsilon_{3,p}=o(p^{-v})$, and $\epsilon_{4,p}=O(s_b  p^{-1})$.
Using the Sherman-Morrison-Woodbury matrix identity, we have
\begin{eqnarray*}
  \bSigma^{-1} &=& (\bB \bSigma_{F} \bB^{\top} + \bSigma_u)^{-1} \cr
  &=& \bSigma_{u}^{-1} - \bSigma_{u}^{-1} \bB (\bSigma_{F}^{-1} + \bB^{\top} \bSigma_{u}^{-1} \bB)^{-1} \bB^{\top} \bSigma_{u}^{-1}
  .
\end{eqnarray*}
Using the Sherman-Morrison-Woodbury matrix identity recursively, we have
\begin{eqnarray*}
  ({\bSigma_{F}}^{-1} + \bB^{\top} \bSigma_{u}^{-1} \bB)^{-1}
  &=&
  \sum_{k=0}^{\infty} (-1)^k ((\bB^{\top} \bSigma_u^{-1} \bB )^{-1} {\bSigma_{F}}^{-1})^k (\bB^{\top} \bSigma_u^{-1}  \bB)^{-1} \cr
  &=& (\bB^{\top} \bSigma_u^{-1}  \bB)^{-1} - (\bB^{\top} \bSigma_u^{-1}  \bB)^{-1} {\bSigma_{F}}^{-1} (\bB^{\top} \bSigma_u^{-1}  \bB)^{-1}  \cr
  &&+  (\bB^{\top} \bSigma_u^{-1}  \bB)^{-1} {\bSigma_{F}}^{-1} (\bB^{\top} \bSigma_u^{-1}  \bB)^{-1} {\bSigma_{F}}^{-1} (\bB^{\top} \bSigma_u^{-1}  \bB)^{-1} - \cdots
  .
\end{eqnarray*}
Let $\mathcal{Y} = (\bB^{\top} \bSigma_u^{-1}  \bB)^{-1} - (\bB^{\top} \bSigma_u^{-1}  \bB)^{-1} {\bSigma_{F}}^{-1} (\bB^{\top} \bSigma_u^{-1}  \bB)^{-1}$.
Then, we have
\begin{eqnarray}\label{eq:woodbury-approx-Z}
  \bSigma^{-1} = \bSigma_{u}^{-1} - \bSigma_{u}^{-1} \bB \mathcal{Y} \bB^{\top} \bSigma_{u}^{-1} + \sum_{k=2}^{\infty} \mathcal{Z}_k
  ,
\end{eqnarray}
where $\mathcal{Z}_{k} = (-1)^k \bSigma_{u}^{-1} \bB   ((\bB^{\top} \bSigma_u^{-1} \bB )^{-1} {\bSigma_{F}}^{-1})^k (\bB^{\top} \bSigma_u^{-1}  \bB)^{-1}  \bB ^{\top} \bSigma_{u}^{-1}$.
Simple algebra shows that there exists a constant $C$ such that $\norm{\mathcal{Z}_{k}}_{\max} = O((\frac{C}{p^{v}})^k)$ for all $k \geq 2$.
Therefore, for sufficiently large $p$, we have
\begin{equation*}%
  \left \lVert \sum_{k=2}^{\infty} \mathcal{Z}_k  \right \rVert _{\max}\leq C p^{-2v}
  .
\end{equation*}
We simply write this as follows:
\begin{equation}\label{eq:negligible-z}
  \sum_{k=2}^{\infty} \mathcal{Z}_k = O(p^{-2v})
  .
\end{equation}
By Lemma \ref{lemma:chi-term} and the fact that
\begin{equation*}
  \left \lVert (\bB^{\top} \bSigma_u^{-1}  \bB)^{-1} {\bSigma_{F}}^{-1} (\bB^{\top} \bSigma_u^{-1}  \bB)^{-1} \right \rVert _{\max} \leq \left \lVert (\bB^{\top} \bSigma_u^{-1}  \bB)^{-1} {\bSigma_{F}}^{-1} (\bB^{\top} \bSigma_u^{-1}  \bB)^{-1} \right \rVert _{2} \leq C p^{-v}
  ,
\end{equation*}
we have
\begin{eqnarray}\label{eq:y-prop}
  [\mathcal{Y}]_{i,j}=
  \begin{cases}
    O(p^{-(1-v)/2} + s_b p^{-v}) , & \text{if } r_c + 1 \leq i,j \leq r \text{ and }  h(i) \neq h(j) ,\cr
    O(1) , & \text{otherwise.}
  \end{cases}
\end{eqnarray}
Let $\mathcal{B} = \bSigma_{u}^{-1} \bB \mathcal{Y} \bB^{\top} \bSigma_{u}^{-1}$.
Using \eqref{eq:woodbury-approx-Z} and \eqref{eq:negligible-z}, we have
\begin{eqnarray}\label{eq:woodbury-approx}
  \bSigma^{-1} &=&  \bSigma_{u}^{-1} - \mathcal{B} + O(p^{-2v}) %
  .
\end{eqnarray}
Then, for $1 \leq i,j \leq p$, we have
\begin{eqnarray}\label{eq:approxB}
  [\mathcal{B}]_{i,j}
  &=& \sum_{l=1}^{p}  \sum_{k=1}^{p}  [\bSigma_u^{-1}]_{i,k} [\bB \mathcal{Y} \bB^{\top}]_{k,l} [\bSigma_u^{-1}]_{l,j} \cr
  &=& \sum_{H\in \mathcal{G}}  \sum_{l \in H}  \sum_{k \in H}  [\bSigma_u^{-1}]_{i,k} [\bB \mathcal{Y} \bB^{\top}]_{k,l} [\bSigma_u^{-1}]_{l,j} \cr
  &&+ \sum_{H_{1}\in \mathcal{G}} \sum_{H_{2}\in \mathcal{G}\setminus \left\lbrace H_{1} \right\rbrace}  \sum_{l \in H_{2}}  \sum_{k \in H_{1}}  [\bSigma_u^{-1}]_{i,k} [\bB \mathcal{Y} \bB^{\top}]_{k,l} [\bSigma_u^{-1}]_{l,j} .
\end{eqnarray}

Consider \eqref{eq:thm1-original-version-within}.
Using \eqref{eq:woodbury-approx}, we have $[\bSigma^{-1}]_{G} = [\bSigma_{u}^{-1}]_G - [\mathcal{B}]_G + O(p^{-2v})$.
There are at most $s_g|G|$ non-zero elements in $[\bSigma_{u}^{-1}]_{G}$ by Assumption \ref{ass:sparsity}.
The last term on the right-hand side of \eqref{eq:woodbury-approx} is $o(p^{-v})$.
Thus, it is sufficient for proving \eqref{eq:thm1-original-version-within} that
\begin{equation}\label{eq:samegroup-density-approx}
  \left|G \times G \right|^{-1}  \sum_{(i,j) \in G \times G} \b1 ([{\mathcal{B}}]_{i,j} \geq \delta_G |G|^{-1} - O(\frac{s_b^2}{p^{(1+v)/2}} + \frac{s_b^3}{p^{2v}} ) ) > \frac{\gamma_{G}^{2}}{s_g^4(r_c+r_G)^{6}} - O(s_b^{2}  |G|^{-2})
  .
\end{equation}
For any $x_1 \in H$, $x_2 \in H'$, and different groups $H \neq H' \in \mathcal{G}$, we have
\begin{eqnarray}\label{eq:BYB}
  \left|[[\bB \mathcal{Y} \bB^{\top}]_{H,H'}]_{x_1,x_2}\right| 
  &=& \left|\bB_{x_1,\cdot} \mathcal{Y} \bB_{x_2,\cdot}\right|  \cr
  &\leq& \left| \sum_{y_1=1}^{r_c}\sum_{y_2=1}^{r_c}  \bB_{x_1,y_1} \mathcal{Y}_{y_1,y_2} \bB_{x_2,y_2} \right|  \cr
  &&+ \left|\sum_{y_1=1}^{r_c} \sum_{y_2 = \bar{r}_{{H'}} - r_{H'} + 1}^{\bar{r}_{{H'}}} \bB_{x_1,y_1} \mathcal{Y}_{y_1,y_2} \bB_{x_2,y_2}\right| \cr
  &&+  \left|\sum_{y_1 = \bar{r}_{H} - r_H + 1}^{\bar{r}_{H}} \sum_{y_2=1}^{r_c} \bB_{x_1,y_1} \mathcal{Y}_{y_1,y_2} \bB_{x_2,y_2}\right| \cr
  &&+  \left|\sum_{y_1 = \bar{r}_{H} - r_H + 1}^{\bar{r}_{H}} \sum_{y_2 = \bar{r}_{{H'}} - r_{H'} + 1}^{\bar{r}_{{H'}}} \bB_{x_1,y_1} \mathcal{Y}_{y_1,y_2} \bB_{x_2,y_2}\right| \cr
  &=& O \left( \frac{1}{p^{(1+v)/2}} + \frac{1}{p} + \frac{s_b}{p^{2v}}  \right) 
  ,
\end{eqnarray}
where the last equality follows from Assumption \ref{ass:sparsity}\ref{sparsity:sparsity} and \eqref{eq:y-prop}.
Using \eqref{eq:approxB} for all $(i,j) \in G \times G$, we have
\begin{eqnarray}\label{eq:same-decomp}
  [\mathcal{B}]_{G} &=&  [\bSigma_u^{-1}]_{G} [\bB \mathcal{Y} \bB^{\top}]_{G} [\bSigma_u^{-1}]_{G} + \sum_{H \in \mathcal{G} \setminus \left\lbrace G \right\rbrace} [\bSigma_u^{-1}]_{G,H} [\bB \mathcal{Y} \bB^{\top}]_{H} [\bSigma_u^{-1}]_{H,G} \cr
  && +   [\bSigma_u^{-1}]_{G} [\bB \mathcal{Y} \bB^{\top}]_{G,G^c} [\bSigma_u^{-1}]_{G^c,G} +   [\bSigma_u^{-1}]_{G,G^c} [\bB \mathcal{Y} \bB^{\top}]_{G^c,G} [\bSigma_u^{-1}]_{G} \cr
  && + \sum_{\substack{H,H' \in \mathcal{G} \setminus \left\lbrace G \right\rbrace \cr H \neq H'}} [\bSigma_u^{-1}]_{G,H} [\bB \mathcal{Y} \bB^{\top}]_{H,H'} [\bSigma_u^{-1}]_{H',G} \cr
  &=&  [\bSigma_u^{-1}]_{G} [\bB \mathcal{Y} \bB^{\top}]_{G} [\bSigma_u^{-1}]_{G} + \sum_{H \in \mathcal{G} \setminus \left\lbrace G \right\rbrace} [\bSigma_u^{-1}]_{G,H} [\bB \mathcal{Y} \bB^{\top}]_{H} [\bSigma_u^{-1}]_{H,G}   \cr
  && +O \left( \frac{s_b^2}{p} + \frac{s_b^2}{p^{(1+v)/2}  }+ \frac{s_b^3}{p^{2v}}    \right)  , 
\end{eqnarray}
where the last equality follows from Assumption \ref{ass:sparsity} and \eqref{eq:BYB}.
By Lemma \ref{lemma:main-term} and Assumption \ref{ass:sparsity}\ref{sparsity:sparsity}, the first term on the right-hand side of \eqref{eq:same-decomp} has at least $\frac{\gamma_G^2 |G|^2}{s_g^{4}(r_c + r_G)^{6}} $ elements that are greater than $\delta_G |G|^{-1}$, where $\gamma_G$ and $\delta_G$ are defined in Lemma \ref{lemma:main-term}.
On the other hand, we have
\begin{eqnarray*}
  && \left\lVert  \sum_{H \in \mathcal{G} \setminus \left\lbrace G \right\rbrace} [\bSigma_u^{-1}]_{G,H} [\bB \mathcal{Y} \bB^{\top}]_{H} [\bSigma_u^{-1}]_{H,G} \right\rVert _{0} \cr
  &\leq& \sum_{H \in \mathcal{G} \setminus \left\lbrace G \right\rbrace} \left\lVert  [\bSigma_u^{-1}]_{G,H} [\bB \mathcal{Y} \bB^{\top}]_{H} [\bSigma_u^{-1}]_{H,G} \right\rVert _{0} \cr
  &\leq& \sum_{H \in \mathcal{G} \setminus \left\lbrace G \right\rbrace} \left\lVert  [\bSigma_u^{-1}]_{G,H} \right\rVert _{0}^2 \cr
  &\leq& s_b^2
  ,
\end{eqnarray*}
where the last inequality is by Assumption \ref{ass:sparsity}\ref{sparsity:sparsity}.
This fact with \eqref{eq:same-decomp} implies \eqref{eq:samegroup-density-approx}, and we conclude that \eqref{eq:thm1-original-version-within} and \eqref{eq:samegroup-density} hold.
Consider \eqref{eq:thm1-original-version-between}.
Using \eqref{eq:woodbury-approx}, we have $[\bSigma^{-1}]_{G,G^c} = [\bSigma_{u}^{-1}]_{G,G^c} - [\mathcal{B}]_{G,G^c} + O(p^{-2v})$.
There are at most $s_b$ non-zero elements in $[\bSigma_{u}^{-1}]_{G,G^c}$ by Assumption \ref{ass:sparsity}.
Thus, it is sufficient for proving \eqref{eq:thm1-original-version-between} that
\begin{equation}\label{eq:diffgroup-density-approx}
  \left|G \times G^c \right|  ^{-1} \sum_{(i,j) \in G \times G^{c}} \b1 \left([|{\mathcal{B}}]_{i,j}| \geq  O \left( \frac{s_b^2}{p^{(1+v)/2}} + \frac{s_b^3}{p^{2v}} \right) \right) < O ( s_b p^{-1})
  .
\end{equation}
Similar to \eqref{eq:same-decomp}, using \eqref{eq:approxB} and Assumption \ref{ass:sparsity}\ref{sparsity:sparsity}, for any $G' \in \mathcal{G} \setminus G$, we have
\begin{eqnarray}\label{eq:diff-decomp}
  [\mathcal{B}]_{G,G'} &=&   [\bSigma_u^{-1}]_{G} [\bB \mathcal{Y} \bB^{\top}]_{G} [\bSigma_u^{-1}]_{G,G'} + [\bSigma_u^{-1}]_{G,G'} [\bB \mathcal{Y} \bB^{\top}]_{G'} [\bSigma_u^{-1}]_{G'}  \cr
  &&+ \sum_{H \in \mathcal{G} \setminus \left\lbrace G,G' \right\rbrace} [\bSigma_u^{-1}]_{G,H} [\bB \mathcal{Y} \bB^{\top}]_{H} [\bSigma_u^{-1}]_{H,G'} \cr
  &&+ [\bSigma_u^{-1}]_{G} [\bB \mathcal{Y} \bB^{\top}]_{G,G'} [\bSigma_u^{-1}]_{G'} + [\bSigma_u^{-1}]_{G,G'} [\bB \mathcal{Y} \bB^{\top}]_{G',G} [\bSigma_u^{-1}]_{G,G'} \cr
  &&+ [\bSigma_u^{-1}]_{G} [\bB \mathcal{Y} \bB^{\top}]_{G,\tilde{G}} [\bSigma_u^{-1}]_{\tilde{G},G'} + [\bSigma_u^{-1}]_{G,\tilde{G}} [\bB \mathcal{Y} \bB^{\top}]_{\tilde{G},G} [\bSigma_u^{-1}]_{G,G'}  \cr
  &&+ [\bSigma_u^{-1}]_{G,G'} [\bB \mathcal{Y} \bB^{\top}]_{G',\tilde{G}} [\bSigma_u^{-1}]_{\tilde{G},G'} + [\bSigma_u^{-1}]_{G,\tilde{G}} [\bB \mathcal{Y} \bB^{\top}]_{\tilde{G},G'} [\bSigma_u^{-1}]_{G'}  \cr
  &&+  \sum_{\substack{H,H' \in \mathcal{G} \setminus \left\lbrace G,G' \right\rbrace  \cr H \neq H'}} [\bSigma_u^{-1}]_{G,H} [\bB \mathcal{Y} \bB^{\top}]_{H,H'} [\bSigma_u^{-1}]_{H',G'} \cr
  &=&  [\bSigma_u^{-1}]_{G} [\bB \mathcal{Y} \bB^{\top}]_{G} [\bSigma_u^{-1}]_{G,G'} + [\bSigma_u^{-1}]_{G,G'} [\bB \mathcal{Y} \bB^{\top}]_{G'} [\bSigma_u^{-1}]_{G'}  \cr
  &&+ \sum_{H \in \mathcal{G} \setminus \left\lbrace G,G' \right\rbrace} [\bSigma_u^{-1}]_{G,H} [\bB \mathcal{Y} \bB^{\top}]_{H} [\bSigma_u^{-1}]_{H,G'} \cr
  &&+  O \left( \frac{s_b^2}{p^{(1+v)/2}} + \frac{s_b^3}{p^{2v}}   \right)  ,
\end{eqnarray}
where $\tilde{G} = \bigcup_{H \in \mathcal{G} \setminus \left\lbrace G, G' \right\rbrace} H$.
Therefore, we can show \eqref{eq:diffgroup-density-approx} by using the fact that
\begin{eqnarray*}
  && \sum_{(i,j) \in G \times G^c} \b1 \left( | [\mathcal{B}]_{i,j} | > O \left( \frac{s_b^2}{p^{(1+v)/2}} + \frac{s_b^3}{p^{2v}} \right) \right)   \cr
  &\leq& \lVert [\bSigma_u^{-1}]_{G} [\bB \mathcal{Y} \bB^{\top}]_{G} [\bSigma_u^{-1}]_{G,G^c} \rVert _{0}  
  + \sum_{G' \in \mathcal{G} \setminus \left\lbrace G \right\rbrace} \lVert [\bSigma_u^{-1}]_{G,G'} [\bB \mathcal{Y} \bB^{\top}]_{G'} [\bSigma_u^{-1}]_{G'} \rVert _{0} \cr
  &&+ \sum_{H \in \mathcal{G} \setminus \left\lbrace G,G' \right\rbrace}  \left \lVert  [\bSigma_u^{-1}]_{G,H} [\bB \mathcal{Y} \bB^{\top}]_{H} [\bSigma_u^{-1}]_{H,G^c} \right \rVert _{0} \cr
  &\leq& |G| \left \lVert [\bSigma_{u}^{-1}]_{G,G^c} \right \rVert _{0}  + \sum_{G' \in \mathcal{G} \setminus \left\lbrace G \right\rbrace} |G'| \left \lVert [\bSigma_{u}^{-1}]_{G,G'} \right \rVert _{0} + \sum_{H \in \mathcal{G} \setminus \left\lbrace G,G' \right\rbrace}  \left \lVert  [\bSigma_u^{-1}]_{G,H} \right \rVert _{0} \left \lVert [\bSigma_u^{-1}]_{H,G^c} \right \rVert _{0}  \cr
  &=& O(s_b p^{v})
  ,
\end{eqnarray*}
where the first inequality is by \eqref{eq:diff-decomp} and the last equality follows from Assumption \ref{ass:sparsity}\ref{sparsity:sparsity}.
Thus, we conclude that \eqref{eq:thm1-original-version-between} and \eqref{eq:diffgroup-density} hold.
$\blacksquare$

\subsection{Proof of Proposition \ref{prop:GLASSO}}
\textbf{Proof of Proposition \ref{prop:GLASSO}.}
We follow the proof of Theorem 1 in \citet{ravikumar2011high}, which applies the primal-dual witness method \citep{wainwright2009sharp}.
The primal-dual witness method constructs both a primal solution and a dual witness that satisfy the optimality conditions simultaneously.
It begins by constructing the primal solution under the assumed sparsity pattern and then checks its asymptotic properties.
It then verifies the solution by checking the sufficient conditions for strict dual feasibility.

We firstly introduce the key terms and notations for the proof.
Let
\begin{align*}
  & \hat{\bOmega}_{E} = \argmin_{{\bOmega} \succ 0,{\bOmega}={\bOmega}^{\top}} \tr (\hat{\bSigma}_E \bOmega) - \log \det \bOmega + \rho_T \norm{\bOmega}_{1} , \cr
  & \tilde{\bOmega}_{E} = \argmin_{{\bOmega} \succ 0,{\bOmega}={\bOmega}^{\top}, \bOmega_{E,S^c}=0} \tr (\hat{\bSigma}_E \bOmega) - \log \det \bOmega + \rho_T \norm{\bOmega}_{1} , \quad \tilde{Z} = \rho_T^{-1} \left( - \hat{\bSigma}_E + \tilde{\bOmega}_E^{-1} \right)  , \cr
  & W = \hat{\bSigma}_{E} - \bSigma_E, \quad \Delta = \tilde{\bOmega}_{E} - \bOmega_E, \text{ and } R(\Delta) = \tilde{\bOmega}_{E}^{-1} - {\bOmega}_{E}^{-1} + {\bOmega}_{E}^{-1} \Delta {\bOmega}_{E}^{-1} 
  ,
\end{align*}
where $S$ is defined in Assumption \ref{ass:GLASSO}(c).
We also define $\bar{A} = \text{vec} (A)$ for the vectorized version of the matrix or set $A$, and $A_\mathcal{T} = (A_{(i,j)})_{(i,j) \in \mathcal{T}}$ for a matrix $A$ and a set of tuples $\mathcal{T}$.
The sub-differential of the norm $\left\lVert \cdot \right\rVert _{1}$ evaluated at some $M \in \mathbb{R}^{p \times p}$ consists of the set of all symmetric matrices $Z \in \mathbb{R}^{p \times p}$ such that
\begin{equation*}
  Z_{ij} =  \begin{cases}
  0 , &  \text{if } i=j ,\cr
  \text{sign} (M_{ij}) , &  \text{if } i \neq j  \text{ and }  M_{ij} \neq 0 ,\cr
  \in [-1,1] , &  \text{if } i \neq j  \text{ and }  M_{ij} = 0 .
  \end{cases}
\end{equation*}

Simple algebra shows that $\kappa_{\Gamma} = \left\lVert \Gamma_{S} ^{-1} \right\rVert _{\infty} = O(1)$, $\kappa_{\Sigma} = \left\lVert \bSigma_E \right\rVert _{\infty} = O(p^v)$, and $\rho_T = O(p^{-4v})$.
By  Lemma 6 in \citet{ravikumar2011high}, we have
\begin{equation}\label{eq:Lemma6-ravi}
  \left\lVert \Delta \right\rVert _{\max} \leq 2 \kappa_{\Gamma} \left( \left\lVert W \right\rVert _{\max} + \rho_T  \right) 
  .
\end{equation}
By (A.2) in Appendix of \citet{choi2023large}, we have
\begin{equation}\label{eq:A2-choi}
  \left\lVert W \right\rVert _{\max} \leq C \left( 1/p^{1-v} + \sqrt{\log p / T} \right) 
  .
\end{equation}
With $v<1/5$, we observe that there exists $T_p = O(p^{8v} \log p)$ such that for $T \geq T_p$,
$$\left\lVert W \right\rVert _{\max} \leq C \left( 1/p^{1-v} + \sqrt{\log p / T} \right) <  \frac{\alpha^2}{24 \kappa_{\Sigma}^{3} \kappa_{\Gamma}^{2}  d (4-2 \alpha)^2}  .$$
Using Lemma 5 in \citet{ravikumar2011high}, we can show
\begin{eqnarray}\label{eq:Lemma5-ravi}
  \left\lVert R(\Delta) \right\rVert _{\max} &<&  \frac{3}{2} d \left\lVert \Delta \right\rVert _{\max} ^2 \kappa_{\Sigma}^{3} \cr
  &\leq& 24 \kappa_{\Sigma}^{3} \kappa_{\Gamma}^{2} \rho_T^{2} d \cr
  &\leq& \frac{\alpha^2}{24 \kappa_{\Sigma}^{3} \kappa_{\Gamma}^{2}  d (4-2 \alpha)^2}
  ,
\end{eqnarray}
where the second inequality is due to \eqref{eq:Lemma6-ravi}.
Therefore, we have
\begin{equation}\label{eq:Lemma4-cond-ravi}
  \max \left\lbrace \left\lVert W \right\rVert _{\max} , \left\lVert R(\Delta) \right\rVert _{\max}  \right\rbrace < \frac{\alpha^2}{24 \kappa_{\Sigma}^{3} \kappa_{\Gamma}^{2}  d (4-2 \alpha)^2}
  .
\end{equation}
The stationary condition for the solution $\tilde{\bOmega}_E$ can be written as
\begin{equation}\label{eq:raw-stationary}
  \hat{\bSigma}_{E} - \tilde{\bOmega}_{E}^{-1}  + \rho_T \tilde{Z}  = 0
  .
\end{equation}
Simple algebra shows that
\begin{eqnarray*} %
  0 &=&  \hat{\bSigma}_{E} - \tilde{\bOmega}_{E}^{-1}  + \rho_T \tilde{Z} \cr
  &=&  \hat{\bSigma}_{E} - \bSigma_{E}  - \left(  \tilde{\bOmega}_{E}^{-1} - \bOmega_{E}^{-1}  + \bOmega_{E}^{-1} \Delta \bOmega_{E}^{-1} \right) + \bOmega_{E}^{-1} \Delta \bOmega_{E}^{-1} + \rho_T \tilde{Z} \cr
  &=& \bOmega_E^{-1} \Delta \bOmega_E^{-1} + W - R(\Delta) + \rho_T \tilde{Z}
  .
\end{eqnarray*}
By vectorizing, the stationary condition can be rewritten as
\begin{equation}\label{eq:stationary}
  \Gamma \bar{\Delta} + \bar{W} - \bar{R}(\Delta) + \rho_T \bar{\tilde{Z}} = 0
  .
\end{equation}
By decomposing \eqref{eq:stationary} into $S$ and $S^c$, and using the fact that $\bar{\Delta}_{S^c} = 0$, we have
\begin{eqnarray}
  && \Gamma_{S} \bar{\Delta}_S + \bar{W}_S - \bar{R}(\Delta)_S + \rho_T \bar{\tilde{Z}}_S = 0  \quad \text{and} \quad  \label{eq:sta-decomp1} \\
  && \Gamma_{S^c,S} \bar{\Delta}_{S} + \bar{W}_{S^c} - \bar{R}(\Delta)_{S^c} + \rho_T \bar{\tilde{Z}}_{S^c} = 0 . \label{eq:sta-decomp2}
\end{eqnarray}
Using \eqref{eq:sta-decomp1} and the fact that $\Gamma_{S}$ is invertible, we have
\begin{equation*}
  \bar{\Delta}_S = (\Gamma_{S})^{-1} (-\bar{W}_S + \bar{R}(\Delta)_S - \rho_T \bar{\tilde{Z}}_S)
  .
\end{equation*}
By substituting this into \eqref{eq:sta-decomp2}, we have
\begin{equation}\label{eq:barZSc}
  \bar{\tilde{Z}}_{S^c} =  \rho_T^{-1} \Gamma_{S^c, S} \left( \Gamma_{S} \right)^{-1} \left( \bar{W}_{S} - \bar{R}(\Delta)_{S} \right) - \rho_T^{-1} \left( \bar{W}_{S^c} - \bar{R}(\Delta)_{S^c} \right) + \Gamma_{S^c , S} \left( \Gamma_{S^c , S} \right)^{-1} \bar{\tilde{Z}}_S 
  .
\end{equation}
Therefore, we have
\begin{eqnarray*}
  \left\lVert \bar{\tilde{Z}}_{S^c} \right\rVert _{\max} &\leq&  \rho_T^{-1} \left\lVert \Gamma_{S^c , S} \left( \Gamma_{S} \right)^{-1} \right\rVert _{1} \left( \left\lVert \bar{W}_{S} \right\rVert _{\max} + \left\lVert \bar{R}(\Delta)_{S} \right\rVert _{\max} \right) \cr
  &&+ \rho_T^{-1}  \left( \left\lVert \bar{W}_{S^c} \right\rVert _{\max} + \left\lVert \bar{R}(\Delta)_{S^c} \right\rVert _{\max} \right) + \left\lVert \Gamma_{S^c , S} \left( \Gamma_{S^c , S} \right)^{-1} \bar{\tilde{Z}}_S \right\rVert _{\max}  \cr
  &\leq& \frac{2-\alpha}{\rho_T} \left( \left\lVert \bar{W} \right\rVert _{\max} + \left\lVert \bar{R}(\Delta) \right\rVert _{\max} \right) + (1-\alpha) \cr
  &<&  1
  ,
\end{eqnarray*}
where the second and last inequalities are due to Assumption \ref{ass:GLASSO}(c) and \eqref{eq:Lemma4-cond-ravi}, respectively.
Therefore, strict dual feasibility is satisfied, and thus $\tilde{\bOmega}_E=\hat{\bOmega}_E$.
We have
\begin{equation}\label{eq:partial-GLASSO-max}
  \left\lVert \hat{\bOmega}_E - {\bOmega}_E \right\rVert _{\max} = \left\lVert \Delta \right\rVert _{\max} \leq 2 C \kappa_{\Gamma} \left( 1/p^{1-v} + \sqrt{\log p / T} + \rho_T  \right) 
  ,
\end{equation}
where the inequality is due to \eqref{eq:Lemma6-ravi} and \eqref{eq:A2-choi}.
Similar to Corollary 3 in \citet{ravikumar2011high}, we can show that
\begin{eqnarray}\label{eq:partial-GLASSO-2}
  \left\lVert \hat{\bOmega}_E - {\bOmega}_E \right\rVert _{2} &\leq&  2 \kappa_{\Gamma} \left( 1/p^{1-v} + \sqrt{\log p / T} + \rho_T  \right)  \min (p^{\frac{1+v}{2}}, p^{v}) \cr
  &=&  O(p^{-3v} + p^{v} \sqrt{\log p / T} )
  .
\end{eqnarray}

Consider $\hat{\bOmega} - {\bOmega}$.
Let $UVU^{\top}$ be the eigendecomposition of $\bB_{c} \bSigma_{c} \bB_{c}^{\top}$, where $U \in \mathbb{R}^{p \times r_c}$ contains the eigenvectors corresponding to the nonzero eigenvalues, and $V \in \mathbb{R}^{r_c \times r_c}$ is the diagonal matrix of these eigenvalues.
By the Sherman-Morrison-Woodbury matrix identity, we have
\begin{equation}\label{eq:GLASSO-decomp}
  \hat{\bOmega} - {\bOmega} = \hat{\bOmega}_E - {\bOmega}_E - R_1
  ,
\end{equation}
where
\begin{eqnarray}\label{eq:R1-decomp}
  R_1 &=& \left( \hat{\bOmega}_{E} - \bOmega_{E} \right)  \hat{U} \left( \hat{V}^{-1} + \hat{U}^{\top} \hat{\bOmega}_{E} \hat{U} \right)^{-1} \hat{U}^{\top} \hat{\bOmega}_{E} \cr
  &&+ {\bOmega}_{E} (\hat{U} - U) \left( \hat{V}^{-1} + \hat{U}^{\top} \hat{\bOmega}_{E} \hat{U} \right)^{-1} \hat{U}^{\top} \hat{\bOmega}_{E} \cr
  &&+ {\bOmega}_{E} {U} \left( \hat{V}^{-1} + \hat{U}^{\top} \hat{\bOmega}_{E} \hat{U} \right)^{-1} (\hat{U}^{\top} - U^{\top}) \hat{\bOmega}_{E} \cr
  &&+ {\bOmega}_{E} {U} \left( \hat{V}^{-1} + \hat{U}^{\top} \hat{\bOmega}_{E} \hat{U} \right)^{-1} {U}^{\top} (\hat{\bOmega}_{E} - {\bOmega}_{E}) \cr
  &&+ {\bOmega}_{E} {U} \left[ \left( \hat{V}^{-1} + \hat{U}^{\top} \hat{\bOmega}_{E} \hat{U} \right)^{-1} - \left( {V}^{-1} + {U}^{\top} {\bOmega}_{E} {U} \right)^{-1} \right]  {U}^{\top} {\bOmega}_{E}  
  .
\end{eqnarray}
Simple algebra shows that
\begin{eqnarray*}
  \lVert \hat{U} - U \rVert _{2}^{2} &\leq& \lVert \hat{U} - U \rVert _{F}^{2} \cr
  &\leq&  p r_c \max_{1\leq i \leq r_c} \lVert \hat{U}_{\cdot, i} - {U}_{\cdot, i} \rVert _{\infty}^{2} \cr
  &\leq& C \left( \frac{1}{p^{2(1-v)}} + \frac{\log p}{T} \right)  
  ,
\end{eqnarray*}
where the last inequality is by Lemmas A.2 and A.3 in Appendix of \citet{choi2023large}.
Let $G_1 = \left( \hat{V}^{-1} + \hat{U}^{\top} \hat{\bOmega}_{E} \hat{U} \right)^{-1}$ and $G_2 = \left( {V}^{-1} + {U}^{\top} {\bOmega}_{E} {U} \right)^{-1}$.
We have
\begin{eqnarray}\label{eq:G1-G2}
  \lVert G_1 - G_2 \rVert _{2} &=& \lVert G_1 (G_1^{-1} - G_2^{-1}) G_2 \rVert _{2} \cr
  &\leq& \lVert G_1 \rVert _{2} \lVert G_2 \rVert _{2} \lVert \hat{V}^{-1} - {V}^{-1} + \hat{U}^{\top} \hat{\bOmega}_{E} \hat{U} - {U}^{\top} {\bOmega}_{E} {U}  \rVert _{2} \cr
  &\leq& C \left( \lVert \hat{V}^{-1} - {V}^{-1} \rVert _{2} + \lVert \hat{U}^{\top} \hat{\bOmega}_{E} \hat{U} - {U}^{\top} {\bOmega}_{E} {U} \rVert _{2} \right) 
  .
\end{eqnarray}
For the first term on the right-hand side of \eqref{eq:G1-G2}, we have
\begin{eqnarray}\label{eq:V1-V2}
  \lVert \hat{V}^{-1} - {V}^{-1} \rVert _{2} &\leq& \lVert \hat{V}^{-1}  \rVert _{2} \lVert {V}^{-1}  \rVert _{2} \lVert \hat{V} - {V} \rVert _{2} \cr
  &\leq& C \frac{1}{p^{2}} \left( p \sqrt{\frac{\log p}{T} } + \lVert \bSigma_{E} \rVert _{2} \right)  \cr
  &\leq& C \left( p^{-1} \sqrt{\frac{\log p}{T} } + p^{-2+v} \right) 
  ,
\end{eqnarray}
where the second inequality is by Lemmas A.1 and A.3 in Appendix of \citet{choi2023large} as well as Weyl's inequality.
For the second term on the right-hand side of \eqref{eq:G1-G2}, we have
\begin{eqnarray}\label{eq:UOU}
  && \lVert \hat{U}^{\top} \hat{\bOmega}_{E} \hat{U} - {U}^{\top} {\bOmega}_{E} {U} \rVert _{2} \cr
  &&\leq \lVert \hat{U}^{\top} \hat{\bOmega}_{E} \hat{U} - \hat{U}^{\top} \hat{\bOmega}_{E} {U} \rVert _{2} + \lVert \hat{U}^{\top} \hat{\bOmega}_{E} {U} - \hat{U}^{\top} {\bOmega}_{E} {U}  \rVert _{2} +  \lVert \hat{U}^{\top} {\bOmega}_{E} {U} - {U}^{\top} {\bOmega}_{E} {U}  \rVert _{2} \cr
  &&\leq C \left( \lVert \hat{U} - U\rVert _{2} + \lVert \hat{\bOmega}_{E} - \bOmega_{E} \rVert _{2} \right) \cr
  &&\leq C \left( p^{-3v} + p^{v} \sqrt{\frac{\log p}{T}}  \right) 
  ,
\end{eqnarray}
where the last inequality is by the fact that $v<\frac{1}{5}$.
Therefore, we can bound the last term on the right-hand side of \eqref{eq:R1-decomp} under the spectral norm as follows:
\begin{eqnarray*}
  && \left \lVert {\bOmega}_{E} {U} \left( G_1 - G_2 \right)  {U}^{\top} {\bOmega}_{E} \right  \rVert _{2} \cr
  &\leq& C \left \lVert  G_1 - G_2 \right \rVert _{2}  \cr
  &\leq& C \left( \lVert \hat{V}^{-1} - {V}^{-1} \rVert _{2} + \lVert \hat{U}^{\top} \hat{\bOmega}_{E} \hat{U} - {U}^{\top} {\bOmega}_{E} {U} \rVert _{2} \right) \cr
  &\leq& C \left( p^{-3v} + p^{v} \sqrt{\frac{\log p}{T} } \right) 
  ,
\end{eqnarray*}
where the second inequality follows from \eqref{eq:G1-G2} and the third inequality follows from \eqref{eq:V1-V2}, \eqref{eq:UOU}, and the fact that $v<1/5$.
Similarly, we can bound the remaining four terms on the right-hand side of \eqref{eq:R1-decomp} under the spectral norm by $C ( p^{-3v} + p^{v} \sqrt{\frac{\log p}{T}})$.
Thus, we conclude
\begin{eqnarray*}
  \lVert \hat{\bOmega} - \bOmega \rVert _{2} &\leq& \lVert \hat{\bOmega}_{E} - \bSigma_{E}^{-1} \rVert _{2} + \lVert R_1 \rVert _{2} \cr
  &\leq&  C \left( p^{-3v} + p^{v} \sqrt{\frac{\log p}{T}}  \right) 
  .
\end{eqnarray*}
$\blacksquare$

\subsection{Proof of Theorem \ref{prop:density-sample}}

\textbf{Proof of Theorem \ref{prop:density-sample}.}
It is sufficient to show that %
\begin{eqnarray}
  && \left|G \times G \right|^{-1}  \sum_{(i,j) \in G \times G} \b1 ([\hat{\bOmega}]_{i,j} \leq - \delta_G |G|^{-1} + \epsilon_{1,p} ) > \frac{\gamma_{G}^{2}}{s_g^4(r_c+r_G)^{6}} - \epsilon_{2,p}  \label{eq:thm2-original-version-within} \\
  &\text{and}& \left|G \times G^c \right|  ^{-1} \sum_{(i,j) \in G \times G^{c}} \b1 (|[\hat{\bOmega}]_{i,j}| < \epsilon_{3,p}) > 1 - \epsilon_{4,p} \label{eq:thm2-original-version-between}
  ,
\end{eqnarray}
where $\epsilon_{1,p}=o(p^{-v})$, $\epsilon_{2,p}=O(s_g p^{-v})$, $\epsilon_{3,p}=o(p^{-v})$, and $\epsilon_{4,p}=O(s_b  p^{-1})$.
Consider indices $i$ and $j$ such that $[\bOmega]_{i,j} \leq - \delta_G |G|^{-1} + \epsilon_{1,p}$. %
By \eqref{eq:thm1-original-version-within}, there exists $\bar{\epsilon}_{1,p}=o(p^{-v})$ such that
\begin{eqnarray*}
  && [\hat{\bOmega}]_{i,j} \leq [\bOmega]_{i,j} + \left\lVert \hat{\bOmega} - \bOmega \right\rVert _{\max} \leq - \delta_G |G|^{-1} + \bar{\epsilon}_{1,p}
  .
\end{eqnarray*}
Therefore, we conclude \eqref{eq:thm2-original-version-within}.

Consider indices $i$ and $j$ such that $|[ \bOmega]_{i,j}| \leq \epsilon_{3,p}$. %
By \eqref{eq:thm1-original-version-between}, there exists $\bar{\epsilon}_{3,p}=o(p^{-v})$ such that
\begin{eqnarray*}
  && \left|[\hat{\bOmega}]_{i,j}\right|  \leq \left|[\bOmega]_{i,j}\right|  + \left\lVert \hat{\bOmega} - \bOmega \right\rVert _{\max} \leq \bar{\epsilon}_{3,p}
  .
\end{eqnarray*}
Therefore, we conclude \eqref{eq:thm2-original-version-between}.
$\blacksquare$

\end{spacing}

\end{document}